\documentclass[12pt,preprint]{aastex}
\usepackage{natbib}

\shorttitle{CHANDRA OBSERVATION OF ABELL 133} 
\shortauthors{FUJITA ET AL.}
\slugcomment{NAOJ-Th-Ap 2002, No.12}

\begin{document}

\title{Chandra Observations of the Disruption
of the Cool Core in Abell~133}

\author{Yutaka Fujita\altaffilmark{1,2},
Craig L. Sarazin\altaffilmark{2},
Joshua C. Kempner\altaffilmark{2},
L. Rudnick\altaffilmark{3},
O. B. Slee\altaffilmark{4}
A. L. Roy\altaffilmark{5},
H. Andernach\altaffilmark{6},
and
M. Ehle\altaffilmark{7,8}}

\altaffiltext{1}{National Astronomical Observatory, Osawa 2-21-1, Mitaka,
Tokyo 181-8588, Japan; yfujita@th.nao.ac.jp}
\altaffiltext{2}{Department of Astronomy, University of 
Virginia, P.O. Box 3818, Charlottesville, VA 22903-0818, USA;
sarazin@virginia.edu,
jck7k@virginia.edu}
\altaffiltext{3}{Department of Astronomy, University of Minnesota, 116
Church Street SE, Minneapolis, MN 55455}
\altaffiltext{4}{Australia Telescope National Facility, CSIRO, PO Box 76,
Epping, NSW 1710, Australia}
\altaffiltext{5}{Max-Planck-Institut f\"ur Radioastronomie, Auf dem
H\"ugel 69, D-53121 Bonn, Germany}
\altaffiltext{6}{Depto.\ de Astronom\'\i a, Univ.\ Guanajuato, Apdo.\
Postal 144, Guanajuato, C.P.\ 36000, GTO, Mexico}
\altaffiltext{7}{XMM-Newton Science Operations Centre, Apartado 50727,
E--28080 Madrid, Spain}
\altaffiltext{8}{Science Operations \& Data Systems Division, Research and
Scientific Support Department of ESA, ESTEC, 2200 AG Noordwijk, The 
Netherlands}

\begin{abstract}
We present the analysis of a {\em Chandra} observation of the galaxy
cluster Abell~133, which has a cooling flow core, a central radio
source, and a diffuse, filamentary radio source which has been
classified as a radio relic.  The X-ray image shows that the core has a
complex structure.  The most prominent feature is a ``tongue" of
emission which extends from the central cD galaxy to the northwest and
partly overlaps the radio relic. Spectral analysis shows that the
emission from the tongue is thermal emission from relatively cool gas at
a temperature of $\sim 1.3$~keV. One possibility is that this tongue is
produced by Kelvin-Helmholtz (KH) instabilities through the interaction
between the cold gas around the cD galaxy and hot intracluster medium.
We estimate the critical velocity and time scale for the KH instability
to be effective for the cold core around the cD galaxy.  We find that
the KH instability can disrupt the cold core if the relative velocity is
$\gtrsim 400\rm\; km\; s^{-1}$.  We compare the results with those of
clusters in which sharp, undisrupted cold fronts have been observed; in
these clusters, the low temperature gas in their central regions has a
more regular distribution. In contrast to Abell~133, these cluster cores
have longer timescales for the disruption of the core by the KH
instability when they are normalized to the timescale of the cD galaxy
motion. Thus, the other cores are less vulnerable to KH
instability. Another possible origin of the tongue is that it is gas
which has been uplifted by a buoyant bubble of nonthermal plasma that we
identify with the observed radio relic.  From the position of the bubble
and the radio estimate of the age of the relic source, we estimate a
velocity of $\sim$700 km s$^{-1}$ for the bubble.  The structure of the
bubble and this velocity are consistent with numerical models for such
buoyant bubbles.  The energy dissipated by the moving bubble may affect
the cooling flow in Abell~133.  The combination of the radio and X-ray
observations of the radio relic suggest that it is a relic radio lobe
formerly energized by the central cD, rather than a merger-shock
generated cluster radio relic.  The lobe may have been displaced from
the central cD galaxy by the motion of the cD galaxy or by the buoyancy
of the lobe.
\end{abstract}


\section{Introduction}
\label{sec:intro}

The advent of the {\it Chandra X-ray Observatory} enables us to study
the intracluster medium (ICM) in clusters of galaxies with superb
spatial resolution.  {\it Chandra} has revealed the dynamics of
substructures in the central region of clusters.  In some clusters,
sharp contact surfaces between the cold gas in merging subcluster cores
and the hot ICM have been observed \citep*{mar00,vik01b,maz01a}.  These
contact surfaces are often called `cold fronts'.  They are interpreted
as cool cores from subclusters which are moving through the main body of
the cluster without losing their identity \citep{mar00,vik01b,maz01a}.
The existence of those cold fronts also provides information about fluid
dynamics, transport properties, and magnetic fields in the ICM.
\citet*{vik01a} argued that magnetic fields provide a surface tension
that stabilizes the cold fronts against the development of
Kelvin-Helmholtz (KH) instabilities.  {\it Chandra} has also provided
more detailed information on cluster merger shocks than the previous
observations with {\it ROSAT} and {\it ASCA}
\citep*[e.g.,][]{hen95,mar99,fur01}.  For Abell~665, \citet{mv01} showed
that radio emission comes from high temperature shock regions; this
suggests that relativistic electrons are accelerated in merger shocks.

Another important result from {\it Chandra} is the discovery of complex
X-ray structures in the central regions of cooling flow clusters with
radio sources.  Recent {\it Chandra} images and earlier results from
{\it ROSAT} show that central radio sources in cooling flow clusters can
inflate bubbles of nonthermal plasma which displace the X-ray emitting
gas \citep*{boe93,car94,hua98,fab00,mcn00,bla01,maz02a}.  The X-ray
images of the central regions of the Hydra~A and Perseus clusters
revealed minima in the X-ray emission coincident with the radio lobes
\citep{mcn00,fab00}.  Subsequently, similar X-ray minima, corresponding
to radio bubbles, were found in Abell~2052 and MKW3s
\citep{bla01,maz02a}; a particularly strong interaction between the ICM
and radio plasma was found in Abell~2052 \citep{bla01}.  For Hydra~A,
\citet{dav01} suggested that the AGN activity may prevent the gas from
cooling to low temperatures \citep[see also][]{ike97}.  This suggests
that these central regions are not simple, spherically-symmetric,
hydrostatic systems.  A wealth of previous X-ray observations has shown
that gas in the cores of many clusters is cooling through at least part
of the X-ray emitting temperature range from $\sim$$10^8$ to
$\sim$$10^7$ K \citep[see][for a review]{fab94}.  Since the cooling time
of gas in many cluster cores is well below the age of the Universe,
cooling flows are the natural consequence of radiative cooling if there
are no balancing heat sources.  More recent spectral observations
suggest that the gas does not continue to cool down to low temperatures
at the rates suggested previously \citep{pet01}.  Previous {\it ASCA}
observations had also suggested this \citep[see][for a review]{mak01}.

Abell~133 is an X-ray luminous cluster at $z=0.0562$ \citep*{way97} with
a central cooling flow \citep*{whi97}.  The central cD galaxy is a radio
source \citep{sle84}.  There also is a diffuse, very steep spectrum
radio source $\sim$30 kpc north of the center of the cD galaxy, which
has no optical counterpart and is classified as a radio relic
\citep{sle01}. The radio relic source has an unusual filamentary
structure.  A comparison of previous {\it ROSAT} X-ray and {\it VLA}
radio observations revealed an excess of X-ray emission near to the
cluster center which was spatially coincident with the radio relic
\citep{riz00,sle01}.  \citet{riz00} argued that the enhanced emission
was the result of the interaction between a jet from the cD nucleus with
hot ICM.  \citet{sle01} indicated that the excess X-ray emission in the
$0.5-2.0$~keV band is too high to be attributed to the inverse Compton
(IC) emission from the high energy particles in the relic alone.  On the
basis of {\it ROSAT} data alone, one cannot draw a conclusion about the
interaction between the radio plasma and the hot ICM because of {\it
ROSAT}'s limited spatial resolution and bandwidth.  In this paper, we
present {\it Chandra} observations of the central region of Abell~133,
which we use to study the X-ray morphology, temperature distribution,
and the contribution of nonthermal emission.  We assume $H_0=70\;\rm
km\;s^{-1}\; Mpc^{-1}$, $\Omega_0=0.3$, and $\Omega_{\Lambda}=0.7$
unless otherwise mentioned.  At a redshift of 0.0562, 1\arcsec\
corresponds to 1.07 kpc.

\section{Spatial Structure}
\label{sec:spatial}

\subsection{X-ray Images}
\label{sec:spatial_images}

Abell~133 was observed with {\it Chandra} on 2000 October 13--14 for a
useful exposure time of 35.5~ks.  The observation was made with the
center of the cluster located between node boundaries near the aim point
on the back-illuminated chip S3.  The detector temperature during the
observation was $-120$ C.  Hot pixels, bad columns, and events with
grades 1, 5, and 7 were excluded from the analysis. A few periods of
time containing background flares were removed 
using the {\sc lc\_clean} software provided by
Maxim Markevitch\footnote{See
\url{http://asc.harvard.edu/cal/Links/Acis/acis/Cal.prods/bkgrnd/current/
index.html}.}.
Only data from the S3 chip are discussed here.

\begin{figure}\epsscale{0.4}
\plotone{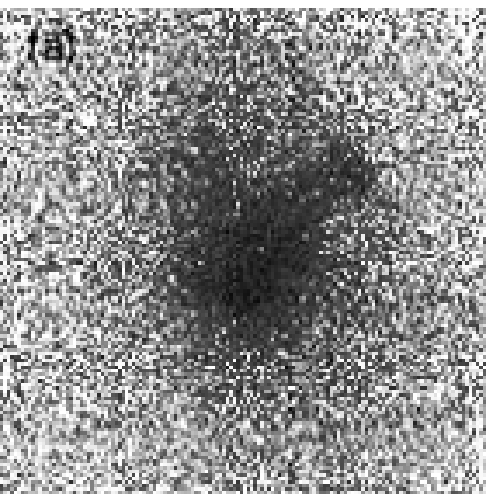} \plotone{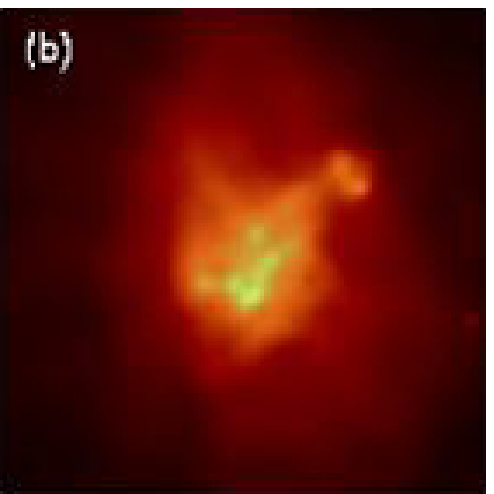} \caption{ (a) Raw {\it Chandra}
X-ray image (0.3--10~keV) of the inner
$2\arcmin\times 2\arcmin$
region of Abell~133, uncorrected for exposure or background.  North is
up and East is left.
The image pixels are 0\farcs492 square.  (b) Adaptively smoothed {\it
Chandra} X-ray image of the same region, corrected for background,
exposure, and vignetting.  \label{fig:csm}}
\end{figure}

Figure~\ref{fig:csm}a shows the raw {\it Chandra} image of the inner
2\arcmin$\times$2\arcmin\ region of the cluster in the $0.3-10$~keV
energy band, uncorrected for exposure or background.
Figure~\ref{fig:csm}b shows an adaptively smoothed image of the same
region.  The {\it Chandra} Interactive Analysis of Observation ({\sc
ciao}\footnote{See \url{http://asc.harvard.edu/ciao/}.} 2.0) routine
{\sc csmooth} was used.  The image has a minimum signal-to-noise ratio
of three per smoothing beam and was corrected for exposure, vignetting,
and background. The background was taken from blank sky observations
compiled by Maxim Markevitch\footnotemark[8].  Complex X-ray structures
are seen within $\sim 1'$ of the center of the cluster.  The most
prominent is a ``tongue''-like feature which extends from the cluster
center to the northwest, and ends in two knots of emission.  This tongue
can be identified with the excess X-ray emission feature observed by
{\it ROSAT} (Fig.~1b in \citealt{riz00} and Fig.~12b in
\citealt{sle01}).  To either side of the tongue, the X-ray surface
brightness is rather low.  At the southeast edge of the faint region to
the east of the tongue, there is a diffuse arc of brighter X-ray
emission which curves toward the northeast. The core region appears to
have a relatively sharp, curved edge to the southeast.

Within the core region, there are several bright knots of X-ray
emission, as well as the two knots at the end of the tongue.  Several of
these knots appear to be small enough to be due to unresolved point
sources.  We used the wavelet source detection method to detect possible
point sources in the image.  The {\sc ciao wavdetect}\footnotemark[9]
routine was used with the significance threshold set at $1 \times 10^{-6}$.
Within the central region of the cluster, three sources were
detected with sizes consistent with point sources; they are indicated in
Figure~\ref{fig:point}.  Source C is the western knot at the tip of the
tongue of X-ray emission.

In Figure~\ref{fig:radio}, the 1.4 GHz radio image from \citet{sle01} is
shown in green, superposed on the adaptively smoothed $Chandra$ X-ray
image (Figure~\ref{fig:csm}b) in red.  Contours from the radio image are
also shown in Figure~\ref{fig:point}.  A fairly compact radio source is
associated with the central cD galaxy.  This appears in yellow in
Figure~\ref{fig:radio} as this is also a peak in the X-ray image.  This
figure also shows the extended filamentary radio relic source to the
NNW.  The bright knots at the tip of the X-ray tongue are projected on
the radio relic source, to the west of the brightest portion, and the
yellow-orange color indicates that this region is bright in both radio
and X-rays.  However, most of the radio relic lies to the east and west
of the tongue, projected on regions of relatively low X-ray surface
brightness.  An interesting aspect of the radio image are the long
filaments extending to the east and south-east of the main portion of
the relic.  These filaments appear to wrap around the ``wing'' of
brighter X-ray emission in this direction.  Also, there is some evidence
that a shorter radio filament wraps around the northern edge of the tip
of the X-ray tongue.

In Figure~\ref{fig:Xopt}, the X-ray contour plot of the central region
of Abell~133 is overlaid on the optical Digital Sky Survey (DSS) image.
As can be seen, there is no optical counterpart of the X-ray tongue,
neither for the entire tongue, nor for the point-like knots (including
source C) at the end of the tongue. X-ray source~B corresponds to a
galaxy (2MASXi J0102379$-$215305); the optical position agrees with the
X-ray position to within 1\arcsec\ and the galaxy is a member of
Abell~133 \citep{way97}. This galaxy is also a compact radio source
\citep{sle01}.  The X-ray source~A is close to the optical center of the
cD galaxy and the core of the central radio source in the cD galaxy, but
is separated by about $3''$ to the south.  It is not clear whether
source~A is really a distinct point source, perhaps due to the radio
active galactic nucleus (AGN) of the cD galaxy, or whether it is just
the peak in the cluster surface brightness profile.  The position of the
cD galaxy, central radio source, and central X-ray peak are all shifted
from the larger scale center (see \S\ref{sec:spatial_profiles}) of the
X-ray emission toward the south.

To help visualize the faint X-ray structures around the cD galaxy, we
compare the observed cluster to a smooth elliptical isophotal model.
Elliptical isophotes are fitted to the image using the IRAF/STSDAS task
ELLIPSE. Before the fitting, we smoothed the image with a Gaussian
kernel with a width of 1\arcsec. Since the X-ray structure around the cD
is very complicated, the ellipticity, position angle, centroid, and
intensity of the elliptical isophotes are all allowed to vary, except
that the centroid of the innermost elliptical isophote is fixed to the
position of the cD galaxy.  The fits to the elliptical isophotes are
used to create a smooth model of the image, which is subtracted from the
original X-ray image of the cluster. The counts of the model image are
multiplied by 0.5 before the subtraction to avoid oversubtraction.
Figure~\ref{fig:residu} shows the subtracted image of the cluster.  The
cluster is nearly axisymmetric about the tongue; faint wings extend from
the center to the NE and SW.  This ``bird-like'' structure and the shift
of the cD galaxy toward the south or southwest may indicate that the cD
galaxy and its environment are moving toward the south or southeast (see
\S~\ref{sec:disc_tongue_KH}).  Note that if we do not multiply by 0.5
before the subtraction, the SW structure becomes unclear because the SW
structure is a ``local'' excess, while on the average, the region to the
north of the cD is brighter than the south in the innermost region
around the cD.

Unlike the other merging galaxy clusters Abell~2142 \citep{mar00},
Abell~3667 \citep{vik01b}, RX~J1720.1+2638 \citep{maz01a}, Abell~2256
\citep{sun02}, MS~1455.0+2232 \citep{maz02c}, and 
1E0657$-$56 \citep{mar02}, Abell~133 does not have clear surface
brightness discontinuities except around the NW tongue and the complex
structure within $\sim 20''$ from the center (Figures~\ref{fig:csm} and
\ref{fig:Xopt}).

\subsection{Radial Profiles of X-ray Surface Brightness and Density}
\label{sec:spatial_profiles}

\begin{figure}\epsscale{0.60}
\plotone{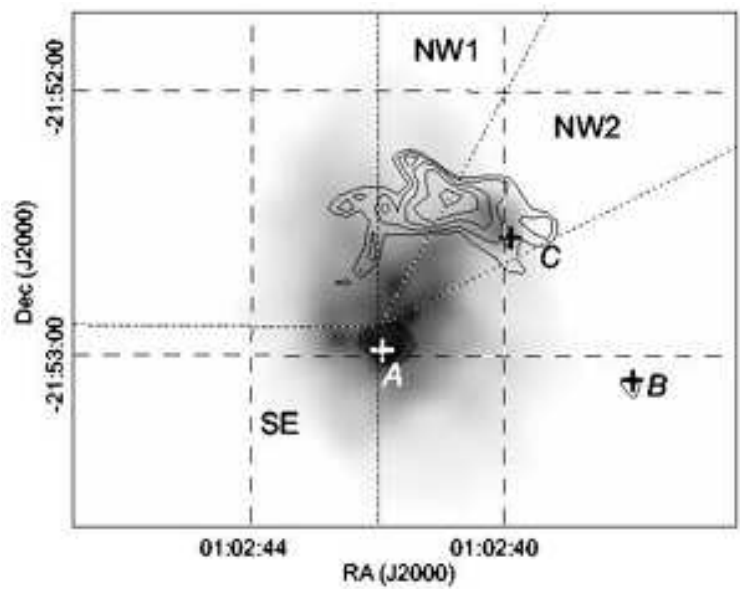} \caption{Radio brightness contours \citep{sle01}
overlaid on the adaptively smoothed image of the central region of
Abell~133; the radio contours are logarithmically spaced by a factor of
$\sqrt{2}$. The positions of several candidate X-ray point sources (A,
B, C) are marked with crosses.  The dotted lines show sectors (SE, NW1,
NW2) used to derive X-ray surface brightness profiles.
\label{fig:point}}
\end{figure}

\begin{figure}\epsscale{0.60}
\plotone{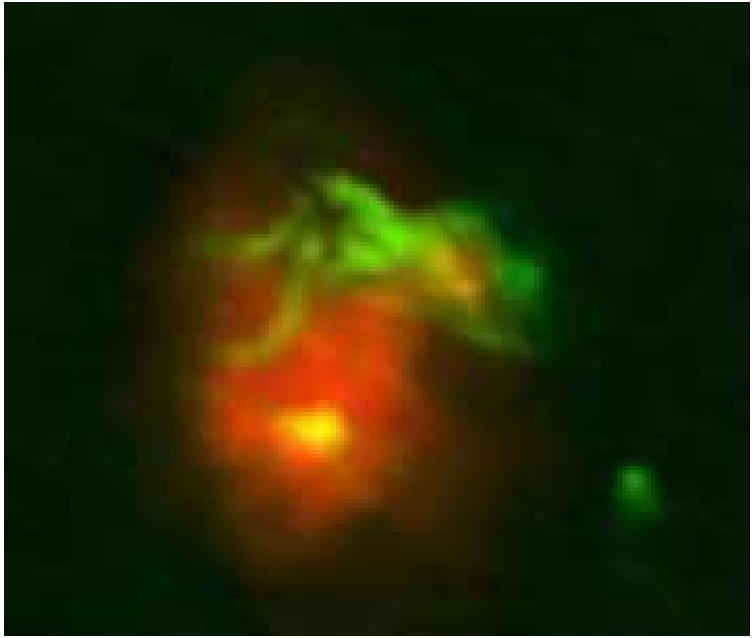} \caption{Radio image at 1.4 GHz \citep{sle01} is
overlaid in green on the adaptively smoothed image of the central region
of Abell~133 (Fig.~\protect\ref{fig:csm}b) in red.
North is up and east is left. The region covered is roughly
2\arcmin$\times$1\farcm7.  \label{fig:radio}}
\end{figure}

\begin{figure}\epsscale{0.60}
\plotone{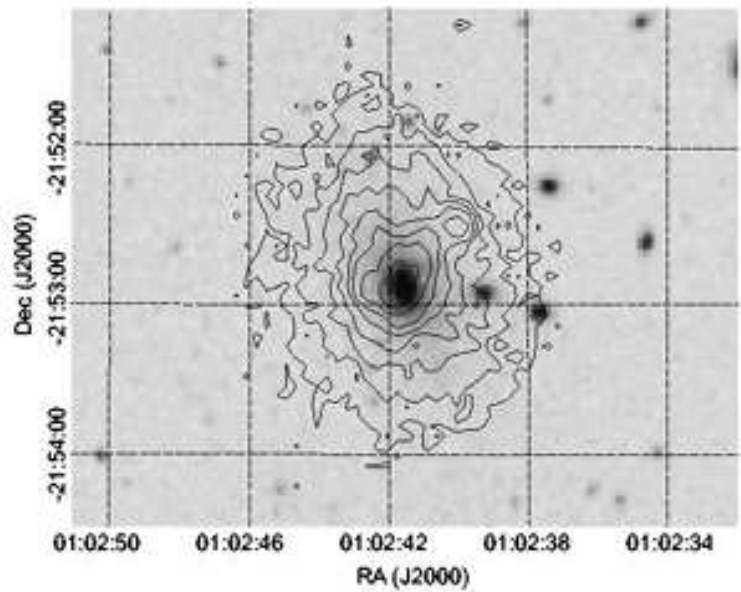}
\caption{X-ray brightness contours ($0.3-10$~keV
band, logarithmically spaced by a factor of $\sqrt{2}$),
overlaid on the DSS
optical image scanned from a SERC-J survey plate taken in 1997 at the UK
Schmidt telescope, Australia.
\label{fig:Xopt}}
\end{figure}

\begin{figure}\epsscale{0.50}
\plotone{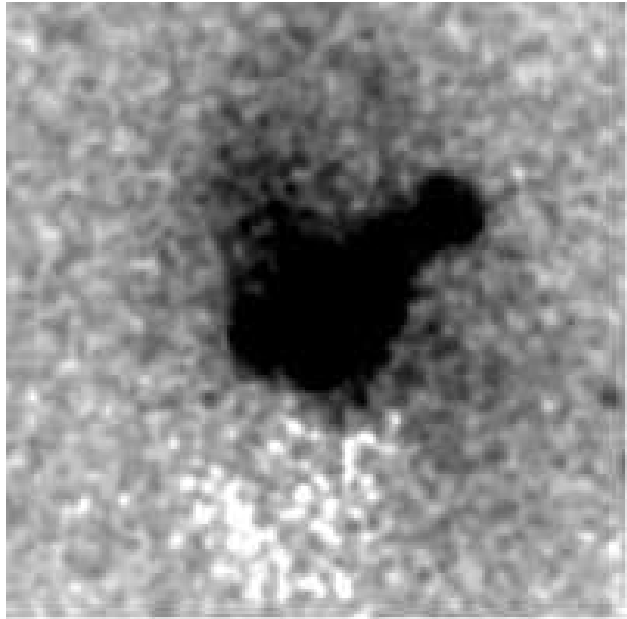} \caption{The residual emission from the central
region of Abell~133 after subtracting a smooth, elliptical isophotal
model. 
The image has been smoothed with a gaussian of $\sigma=1''$. 
North is up and east is left. The image covers $2\arcmin\times
2\arcmin$.  \label{fig:residu}}
\end{figure}

\begin{figure}\epsscale{0.50}
\plotone{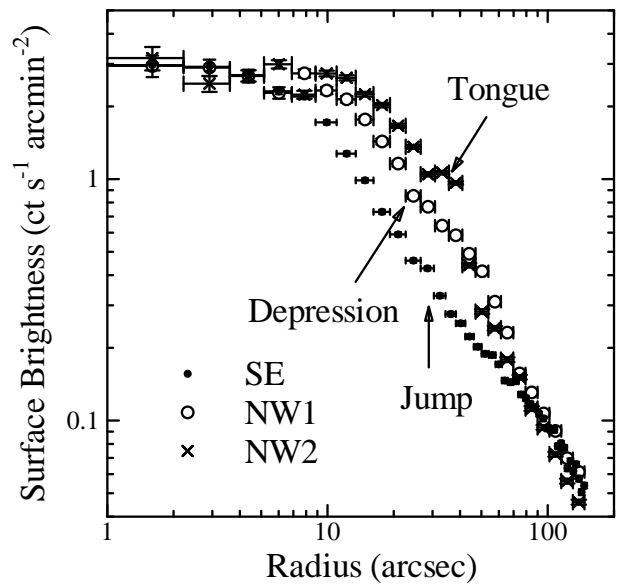} \caption{Surface brightness as a function of radius
 for the three sectors shown by dotted lines in Figure~\ref{fig:point}.
Error bars are 1-$\sigma$ Poisson uncertainties,
but they are small and difficult to see except in the inner region.
\label{fig:surf}}
\end{figure}

In order to study the X-ray structure quantitatively, we extracted
surface brightness profiles for the $0.3-10$~keV band for three sectors
shown in Figure~\ref{fig:point}.  The resulting surface brightness
profiles are shown in Figure~\ref{fig:surf}.  They were derived by
accumulating counts in circular annular wedges.  The range of radii
covered by each annulus is shown by the horizontal bars attached to each
point.  There are vertical error bars, but they are too small to be
easily seen in the Figure.  Because of the complexity of the structure
in the central regions, it is difficult to select a center for the
sectors.  The center that we selected ($01^{\rm h}02^{\rm m}42\fs1$;
$-21\degr52\arcmin54\arcsec$, J2000) is the centroid of the bright core
of the X-ray emission, which we took to be outlined by the contour
having a surface brightness of 4 counts/pixel (0.3--10 keV).  As is
obvious from Figure~\ref{fig:point}, the
sector center does not coincide with the position of the cD galaxy
($01^{\rm h}02^{\rm m}41\fs7$; $-21\degr52\arcmin56\arcsec$, J2000), nor
does it lie on any of the brighter knots of X-ray emission within the
core.  However, it does lie along the extension of the X-ray tongue into
the center of the core, which makes it easier to isolate the surface
brightness profile of the tongue (NW2 sector).  With this center, the
NW1 profile passes through much of the radio-emitting area outside of
the region of the tongue.  Also, the drop in the surface brightness
associated with the edge of the bright core region
(Figures~\ref{fig:csm}, \ref{fig:Xopt}, and~\ref{fig:residu}) can be
studied with this center.

The SE profile shows that there is a small jump corresponding to a
possible bow-shaped feature about 30\arcsec\ from the central cD
(Figures~\ref{fig:csm}, \ref{fig:Xopt}, \ref{fig:residu}, and
\ref{fig:surf}).
While the SE profile is relatively smooth, the NW1 profile shows a
depression at $r\sim 25''$ relative to a smooth interpolation of the
declining surface brightness from smaller to larger radii.  The position
of this depression corresponds to that of the radio relic
(Figure~\ref{fig:point}).  The NW1 sector also features a high surface
brightness region associated with one of the X-ray knots near the center
of the cD (at 8\arcsec).  The NW2 profile shows a hump or plateau at
about the same radius of the depression, followed by a very rapid drop
in surface brightness at about 40\arcsec\ (Figure~\ref{fig:surf}).  The
hump represents the tip of the X-ray tongue.

For the SE sector, we also determined the electron density profile by
deprojecting the surface brightness profiles in Figure~\ref{fig:surf}
assuming that the density distributions are spherically symmetric and
the density is constant in spherical shells
(Figure~\ref{fig:density}). In the SE sector, the density declines
smoothly except for a small jump at a radius of about 30\arcsec, which
corresponds to the small jump in Figure~\ref{fig:surf}. For the NW
sectors, we show the density profiles outside the tongue and the relic
because the assumption of spherical symmetry is not correct in the
central regions.

\section{Spectral Analysis}
\label{sec:spectra}

Although we analyze the spectra in detail for a number of regions, a
sense of the overall temperature distribution of the cluster can be
found from an X-ray color image. For three energy bands ($0.3-1.5$,
$1.5-2.5$, and $2.5-10$~keV), we make adaptively smoothed images as in
Figure~\ref{fig:csm}b. We allocate the colors of red, green, and blue to
the $0.3-1.5$, $1.5-2.5$, and $2.5-10$~keV band images, respectively,
and combine the three images. Figure~\ref{fig:color} shows the result;
the cluster center and the tongue are red, implying that these contain
cooler gas.
In particular, the tip of the tongue appears to be the
coolest. There are no strong, sharp features which are blue (hard
spectra due to hot gas) and which might be attributed to shocks.

We extracted spectra in the $0.7-10$~keV band in PI channels from
selected regions of the {\it Chandra} image using the {\sc
ciao}\footnotemark[9] software package.  We used the gain file
acisD2000-01-29gainN0001.fits and the fef file
acisD2000-01-29fef-piN0001.fits.  Both the response matrix files and the
ancillary response files were calculated using the {\sc
calcrmf/calcarf}\footnote{See
\url{http://asc.harvard.edu/cont-soft/software/}.}  1.07 package written
by Alexey Vikhlinin and Jonathan McDowell. The package weighted the
response files by the X-ray brightness over the corresponding image
region.  The spectra were grouped to have a minimum of 20 counts per
bin, and fitted with one or two thermal models
\citep*[MEKAL,][]{kaa93,lie95} using {\sc xspec 11.00}. In general, we
allowed the absorbing column to vary from the Galactic value.  In some
regions, we fitted the spectra with a power-law model.  Errors on fitted
spectral parameters are given at the 90\% confidence level.

\subsection{Average Spectrum}
\label{sec:spectra_avg}

First, we studied the average spectrum extracted from the innermost
circular region of 2\farcm5 radius. We found that this spectrum cannot
be reproduced by a single thermal component (Table~\ref{tab:tot}),
because this cluster has a temperature gradient
(\S~\ref{sec:spectra_rad}).  The temperature at the cluster center is
much lower than that in the outer region, so we fitted the spectrum with
two thermal components with the same absorption.  We assumed that the
metallicities of the two thermal components were the same, because we
could not constrain the metal abundance when they were fitted
independently. We show the result in Table~\ref{tab:tot}. For
comparison, we present the previous results on Abell~133 from the {\it
Einstein} and {\it EXOSAT} observatories \citep{dav93,edg91}.  The
higher of the two temperatures obtained with {\it Chandra} is consistent
with the temperatures obtained from {\it Einstein} and {\it EXOSAT}.
Because of their relatively large fields, {\it Einstein} and {\it
EXOSAT} appear to be affected by the temperature of the outer region.
The {\it Chandra} flux is smaller than that measured previously because
our spectrum applies only to the central portion of the cluster.  The
{\it Chandra} hydrogen column density is higher than the Galactic value
\citep[$1.58\times 10^{20}\rm\; cm^{-2}$;][]{sta92}.

\subsection{Radial Gradients in the Spectra}
\label{sec:spectra_rad}

Figure~\ref{fig:temp} shows the radial profiles of temperature for four
sectors. The position angles are $0^{\circ}-90^{\circ}$ (NE),
$90^{\circ}-180^{\circ}$ (SE), $180^{\circ}-270^{\circ}$ (SW), and
$270^{\circ}-360^{\circ}$ (NW), where the angles are measured from north
through the east.  The center of the sectors is the same as that of the
sectors in Figure~\ref{fig:point}.  For the NW sector, we derive the
temperature of the X-ray tongue separately from that of the surrounding
emission.  The spectra were fitted by a MEKAL model with variable
absorption.  For most regions, the single temperature model can
reproduce the spectra ($\chi^2$/dof $\lesssim 1.3$), where dof is the
degrees of freedom, except for the region of the tongue ($\chi^2$/dof $=
1.58$).  In Figure~\ref{fig:temp}, the temperature increases gradually
outward except for the NW sector.  For the NW sector, the temperature
shows a sudden rise with radius at $r\sim 20''$ for the profile
excluding the tongue.  A radius of $r\sim 20''$ corresponds to the
southern edge of the deficit of X-ray emission associated with the radio
relic (Figure~\ref{fig:surf}).  The fitted absorbing column density in
the central regions is $N_{\rm H}\lesssim 1\times 10^{21}\rm\; cm^{-2}$.
The profile is not inconsistent with a uniform absorbing column.  The
temperature profiles in Figure~\ref{fig:temp} do not change
significantly even if we fix the absorbing column density at the
Galactic value \citep[$1.58\times 10^{20}\rm\; cm^{-2}$;][]{sta92}.

\subsection{Spectra for X-ray Tongue and Radio Relic}
\label{sec:spectra_reg}

We investigated more closely the X-ray spectra of the regions containing
the radio relic and the tongue.  The regions are shown in
Figure~\ref{fig:relic}.  In the radio relic, our main purpose was to
detect or limit the contribution of nonthermal inverse Compton (IC)
emission to the spectrum.  We first fitted each region's spectrum using
a single temperature model with a variable absorbing column density
$N_{\rm H}$.  This model is referred to as ``1T''.  The free parameters
in this model are the temperature, $T$, the metallicity of the gas, $Z$,
measured relative to the solar photosphere values of \citet{and89}, the
absorbing column density, $N_{\rm H}$, and the normalization, $K$.  In
the ``1TPL'' model, we added a power-law component to the 1T model.
Thus, the normalization and the power-law photon index ($\Gamma$) were
added to the free parameters.  On the other hand, we fixed the
metallicity in the 1TPL model at the value of the 1T model, because it
could not be constrained.  Given the radial temperature gradient
(Figure~\ref{fig:temp}), spectral complexity may also result from
several different emission temperatures along the line of sight. Thus,
we also tried a two-temperature model ``2T'' with a variable absorption.
We assume that metallicities of the two thermal components are the same.

For the radio relic region, the model with a power-law nonthermal
component (1TPL) is slightly favored (Table~\ref{tab:region} and
Figure~\ref{fig:sp}a).  However, the difference in $\chi^2$ is not very
significant.  Given the likely calibration uncertainties, we do not
regard the detection of a power-law component as established.  Instead,
we take the upper limit on the flux of this component as an upper limit
on any power-law nonthermal spectral component.  The upper limit on the
$0.3-10$~keV flux of the power-law component is $7.1\times 10^{-13}\;\rm
erg\; cm^{-2}\; s^{-1}$. We discuss the implications of this in
\S\ref{sec:disc_relic_B}.

The spectrum of the X-ray tongue is not fitted very well by a single
temperature model, but is fitted adequately by either the 2T or 1TPL
model.  The 2T model is a somewhat better fit.  In the 1TPL model, the
fraction of the flux in the power-law component is relatively small
(24\% of the 0.3--10 keV flux), so most of the emission is thermal and
relatively cool in any case.  We note that the emission from the tongue
can be explained by the superposition in projection of cold thermal
emission from the small volume of the tongue, and the ambient hotter
thermal emission in the cluster in front of and behind the tongue.  This
would account naturally for the need for two temperature components.  To
test this idea, we tried a model in which we fixed the absorbing column,
the temperature, and the metallicity of the hotter gas spectral
component to those derived from the spectrum of the region surrounding
the tongue.  We also fix the normalization of the hotter component to
the expected value given the foreground and background volume of the
tongue region.  This model is listed as ``2T$'$'' in
Table~\ref{tab:region}.  The f-test indicates that this constrained
model is as good a fit as the unconstrained two temperature model.  We
find that the intrinsic temperature of the cold gas in the tongue is
1.3~keV in the 2T$'$ model.  In this model, the metallicity of the
cooler component in the tongue is allowed to differ from that of the
surrounding gas; the best-fit value is slightly lower but agrees within
the errors.  Nonthermal emission from the tongue is not required
(Table~\ref{tab:region} and Figure~\ref{fig:sp}b).  In this projection
model, we assume that the tip of the tongue is a sphere with a radius of
8~kpc $= 7.5$\arcsec.  The required electron density in the tongue is
approximately $0.03\rm\; cm^{-3}$, which implies that the tongue is
nearly in pressure equilibrium with the ambient medium at the same
radius, whose pressure is $0.026$ keV cm$^{-3}$ $= 4.2 \times 10^{-11}$
dyn cm$^{-2}$.

While there is a peak in the X-ray surface brightness at a position
that nearly coincides with the center of the cD galaxy which is the
host of a radio source, it is uncertain whether this is a peak in the
density of the diffuse X-ray gas or X-ray emission from an AGN.  We
determined the X-ray spectrum from a circular region with a radius of
5\arcsec\ centered on the X-ray brightness peak.  The emission appears
to be purely thermal, and does not show any evidence for a nonthermal,
power-law component.  The spectrum was consistent with an absorbing
column that is identical to that for the surrounding region.  These
both suggest that the peak is emission from diffuse hot gas, and not
primarily due to an AGN.

\subsection{Cooling Flow Spectrum}
\label{sec:spectra_cool}

In the central regions of the Abell~133 cluster, the gas densities are
high and the radiative cooling times are short.  Thus, we fitted the
spectra of the inner regions using a cooling flow model.
For ease of comparison with Figure~\ref{fig:temp}, the spectra
were accumulated in a number of annuli, centered on the center of the
sectors in Figure~\ref{fig:point}.
The radii of the three annuli were selected to
cover the central bright region except the tongue, the region containing
the tongue, and the largest outside region fitting completely on the S3
chip.  As a spectral model, we use the MKCFLOW model based on the MEKAL
plasma emission code.  We include any intrinsic absorption, $\Delta N_{\rm
H}$, that may be acting on the cooling flow in the core of the cluster
using the ZWABS model.  Moreover, we add a MEKAL model representing the
emission of the ICM outside the cooling flow, subject to Galactic
absorption using the WABS model.  We fix the Galactic absorption at
$1.58\times 10^{20}\rm\; cm^{-2}$ \citep{sta92}. The combination of the
models is similar to that of \citet*{sch01}:
\begin{eqnarray}
 {\rm Model_{cf}} &=& 
{\rm WABS_{Gal}}\times [{\rm MEKAL}(T_{\rm High};Z;K) \nonumber \\
 & & +\; {\rm ZWABS}(\Delta N_{\rm H})\times {\rm MKCFLOW}(\dot{M})]
\:.
\end{eqnarray}
The free parameters are given in parentheses. The intrinsic absorption
$\Delta N_{\rm H}$ acts only on the cooling flow emission. The
normalization of the MKCFLOW model, $\dot{M}$, is the so-called cooling
rate or mass deposition rate of the flow. We fix the metallicity $Z$ and
initial gas temperature $T_{\rm High}$ in the MKCFLOW component to the
values of metallicity and temperature of the MEKAL component,
respectively. The results are shown in Table~\ref{tab:cool}. The total
mass deposition rate, $\dot{M}(<r)$, increases outward.  We note that
the mass deposition rate of the whole region ($86^{+26}_{-24}\;
M_{\sun}\;\rm yr^{-1}$; $r<161$~kpc) is somewhat larger than the
deposition rate of $56^{+36}_{-34}\; M_{\sun}\rm\; yr^{-1}$ (corrected
to $H_0=70\;\rm km\;s^{-1}\; Mpc^{-1}$) from \citet{whi97}, although the
errors overlap.  For the entire region, the luminosity of the cooling
flow component is $2.5\times 10^{43}\rm\; erg\; s^{-1}$.

We note that if we do not use the data below 0.9~keV, $\Delta N_{\rm H}$
is significantly reduced (parenthesis in Table~\ref{tab:cool}), and at the
same time, $\dot{M}$ also decreases. Both $\Delta N_{\rm H}$ and $\dot{M}$
strongly depend on the data below 0.9~keV and they correlate with one
another. This effect suggests that there may still be some calibration
uncertainties for {\it Chandra} below 0.9~keV. Thus, we regard the
evidence for excess absorption as preliminary.

\section{Discussion}
\label{sec:disc}

\subsection{The Origin of the X-ray Tongue}
\label{sec:disc_tongue}

The most prominent feature of the {\it Chandra} image of Abell~133 is
the tongue of colder X-ray emitting gas that extends from the center.
Cold filaments similar to the tongue are seen in A1795 and MKW3s with
{\it Chandra} \citep{fab01,maz02a} and in Virgo with {\it XMM-Newton}
(\citealt{bel01}) and {\it ROSAT} (\citealt{boe95}).  We discuss several
possible explanations for the tongue.  The four panels in
Figure~\ref{fig:ori} illustrate these possibilities schematically.

\subsubsection{A Cooling Wake}
\label{sec:disc_tongue_wake}

One possible origin of the tongue is a cooling wake
(Figure~\ref{fig:ori}a).  That is, the tongue might be gas which is
cooling from the hot ICM, attracted into a wake along the path of the
moving cD galaxy \citep{dav94,fab01}. However, the temperature
discontinuity between the tongue and the ambient hot gas is very sharp.
We extracted the X-ray spectra of $5''$ wide regions along but just
outside both sides of the tongue and find the temperature is
$2.7_{-0.5}^{+0.5}$~keV, which is significantly higher than that of the
tongue ($1.7_{-0.1}^{+0.0}$~keV).  This shows that the temperature of
the gas changes on a scale of $\lesssim 5$~kpc.  If the cooling wake
scenario is correct, we expect a smooth temperature profile around the
tongue.  Thus, our result does not seem to support the idea that the
tongue is the cooling gas accreted from the ambient hot gas.

\subsubsection{Convection by a Cluster Merger}
\label{sec:disc_tongue_convection}

Recently, numerical simulations of cluster mergers by \citet{ric01}
showed that ram pressure induced by a cluster merger can displace the
core gas from the cluster's potential center.  As a result, the gas
becomes convectively unstable, and in some cases, a convective plume of
low entropy gas is formed.  In the region of this plume, the variations
in entropy of the gas are much greater than the variations in the
pressure; thus, the low entropy plume has a high density, a low
temperature, and a very high X-ray emissivity.  The plume and wings with
low entropy gas shown at $t=5.75$ and 6.00~Gyr in Figure~14 of
\citet{ric01} are remarkably similar to the X-ray distribution of
Abell~133 (Figures~\ref{fig:csm} and~\ref{fig:residu}).  Thus, based on
the X-ray image alone, merger-induced convection would be a strong
candidate to explain the origin of the X-ray tongue and overall X-ray
morphology of the core of Abell~133.

However, other observations are not fully consistent with this scenario,
at least if the simulations in \citet{ric01} are typical.  In this
mechanism, the plume of dense gas must be falling towards the center of
the cluster potential, and Figure~14 in \citet{ric01} shows that the tip
of the plume should be at the center of the cluster potential.  Since
the stars in the cD galaxy will not be displaced by ram pressure, the
optical center of the cD galaxy should be near to the center of the cluster
potential.  However, the {\it Chandra} image shows that the cD galaxy is
not at the tip of the X-ray tongue (Figure~\ref{fig:point}).  It is
possible that the cD galaxy has orbital motion which results in a
separation from the center of the cluster potential.  In support of this
idea is the observation that the cD is displaced from the center of the
cluster determined on larger scales from the X-ray image
(\S~\ref{sec:spatial_profiles}).

The time when the convective plume is seen corresponds to the time
between the $(b=2 r_s)$, $+$2 Gyr and $+$3.5 Gyr panels in Figures 6
and~7 of \citet{ric01}.  At this time, a smaller merging subcluster
should be located $\sim 0.5$ Mpc to the northwest of the main cluster
center.  Merger shocks would be located to the southeast ($\sim 1$ Mpc)
and to the northwest near the smaller subcluster.  There is no clear
optical evidence for a concentration of galaxies or a cD galaxy to the
northwest where the subcluster should be located.  Moreover, an X-ray
image obtained with {\it ROSAT}, which has a larger field of view than
{\it Chandra}, showed that there is no significant substructure at
larger radii \citep{riz00}.

If this merger scenario can be applied to Abell~133, the radio relic may
be induced by a shock associated with the same merger.  This would
probably require that the merger be at a slightly later stage than
suggested by the \citet{ric01} simulations, and that the smaller
subcluster (located to the northwest) is now falling back into the main
cluster (just after the $[b=2 r_s]$, $+$3.5 Gyr panels in Figures 6 and
7 of \citealp*{ric01}).  A weak bow shock would be formed ahead of this
subcluster; this shock would now be located between the subcluster and
the cD galaxy of Abell~133, and might be accelerating particles to form
the observed radio relic.  It would be a coincidence that the tip of the
X-ray tongue and the radio relic overlapped. Moreover, we do not find a
temperature jump corresponding to a shock at the position of the radio
relic.  Below, we argue more generally that the radio relic is not
produced by a merger shock (\S~\ref{sec:disc_relic_origin}).

\subsubsection{Kelvin-Helmholtz Instability}
\label{sec:disc_tongue_KH}

The complicated X-ray structures in the inner region of the cluster
($r\lesssim 20''$) and in the tongue suggest that some instability is
developing at the cluster center. In this subsection, we show that the
Kelvin-Helmholtz (KH) instability is a possible candidate. In fact, the
shift of the cD galaxy from the X-ray geometrical center to the SE and
the bird-like feature in Figure~\ref{fig:residu} may suggest that the
galaxy is moving toward the SE.

In contrast with Abell~133, distinct smooth discontinuities of surface
brightness or cold fronts have been found for Abell~2142, Abell~3667,
RX~J1720.1+2638, A2256, and MS~1455.0+2232
\citep{mar00,vik01b,maz01a,sun02,maz02c}. For these clusters, the cold
front is interpreted as the boundary between cold gas around the central
cD galaxies and the ambient hot gas. From the shape of the cold front,
the authors concluded that the cD and any associated subcluster are
moving in the direction of the cold front.
\citet{vik01a} indicated that the KH instabilities on small scales must
be suppressed over an extended region (angular width $\sim$30$^\circ$)
near the stagnation point at the cold front for Abell~3667, and that
magnetic fields are likely to be responsible for this suppression.  On
larger scales, the relatively smooth distributions of the cold gas at
the forward edge of these cold fronts suggest that the KH instability on
the scale of the entire cold core
is also suppressed or has not developed to the level that it
significantly changes the core shape for these clusters.

When a gravitational field exists, perturbations at a sharp interface
with wavenumber less than $k$ are stable if
\begin{equation}
\label{eq:KH}
 g\gtrsim \frac{\rho_1 \rho_2 k U^2}{\rho_2^2-\rho_1^2}
=\frac{D k U^2}{D^2-1}
\:,
\end{equation}
where $g$ is the gravitational acceleration at the interface between a
high density region with density $\rho_2$ which is moving with a
relative velocity $U$ through a surrounding medium with a lower density
$\rho_1$ \citep{pat83}, and the density contrast is $D \equiv
\rho_2/\rho_1$.  If the high density gas is approximately isothermal and
hydrostatic in the gravitational field of the cD galaxy and/or an
associated subcluster, and the gravitational field is approximated by an
isothermal sphere, then the gravitational acceleration is roughly
\begin{equation}
\label{eq:grav}
 g \approx \frac{2 k_B T_2}{\mu m_H r_2}\:,
\end{equation}
where $k_B$ is the Boltzmann constant, $T_2$ is the temperature of the
high density region, $\mu$ is the mean molecular weight (0.61), $m_H$ is
the hydrogen mass, and $r_2$ is the radius of the high density region.
We note that equation~(\ref{eq:grav}) may somewhat overestimate the
gravitational acceleration.  This can occur if the outer part of a
subcluster is stripped by tidal forces, and only the core with a flat
density distribution remains.  On the other hand, if the center of the
subcluster has a density cusp rather than a core,
equation~(\ref{eq:grav}) will give a more accurate estimate
\citep[e.g. Fig.~6 in][]{kly99}.

We are interested in the largest scale KH instabilities
with $k \sim k_0 \equiv 2\pi/r_2$
which would lead to a significant disruption of the entire cold front.
Then, equations~(\ref{eq:KH}) and (\ref{eq:grav})
can be rewritten to show that the gas will be stable if
\begin{equation}
\label{eq:KHU}
 U \lesssim U_{\rm KH} \equiv
\sqrt{\frac{k_B T_2}{\pi \mu m_H} \frac{D^2-1}{D}} \, .
\end{equation}
On the other hand, if gravity is unimportant, the growth rate in the
linear regime at a flat interface between two incompressible fluids with
different densities and relative motion is given by
\begin{equation}
\label{eq:KHw}
 \omega = k \frac{D^{1/2}U}{1+D}\:,
\end{equation}
\citep{dra81}.  Here, we set $k \sim k_0 = 2\pi/r_2$.  \citet{mur93}
showed that conditions (\ref{eq:KHU}) and (\ref{eq:KHw}) are consistent
with results of numerical simulations.

\begin{figure}\epsscale{0.50}
\plotone{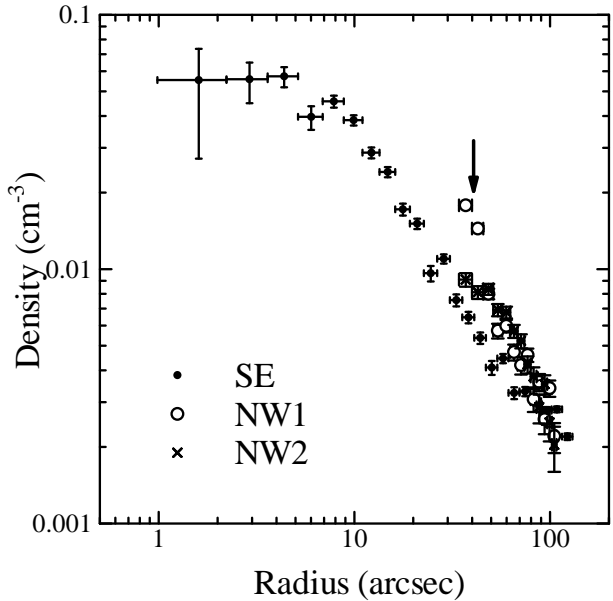} \caption{Electron density as a function of radius
 for the sectors shown by dotted lines in Figure~\ref{fig:point}. For
 NW1 and NW2 sectors, we excluded the region of the tip of the
 tongue 
(the radius of which is indicated by the arrow).
Error bars are 1-$\sigma$.
\label{fig:density}}
\end{figure}

\begin{figure}\epsscale{0.45}
\plotone{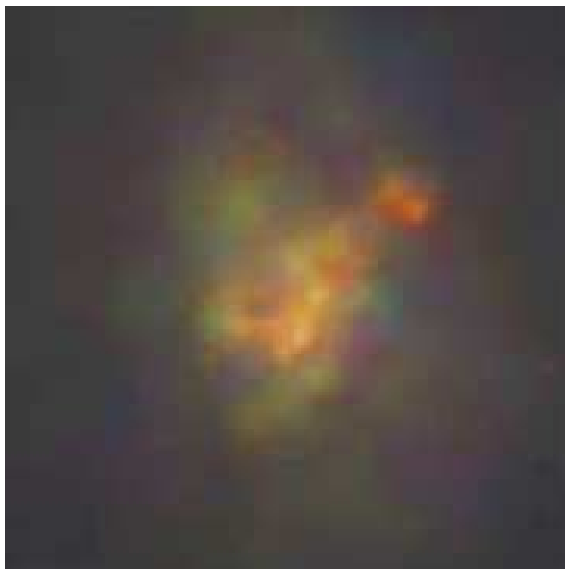} \caption{X-ray color image of the central region of
Abell~133, produced by assigning red, green, and blue to the $0.3-1.5$,
$1.5-2.5$, and $2.5-10$~keV band, respectively.
Thus, red regions have unusually soft X-ray emission, while green-blue
regions have unusually hard emission.
North is up and east is left.
The image covers 2\arcmin $\times$2\arcmin.
\label{fig:color}}
\end{figure}

\begin{figure}\epsscale{0.50}
\plotone{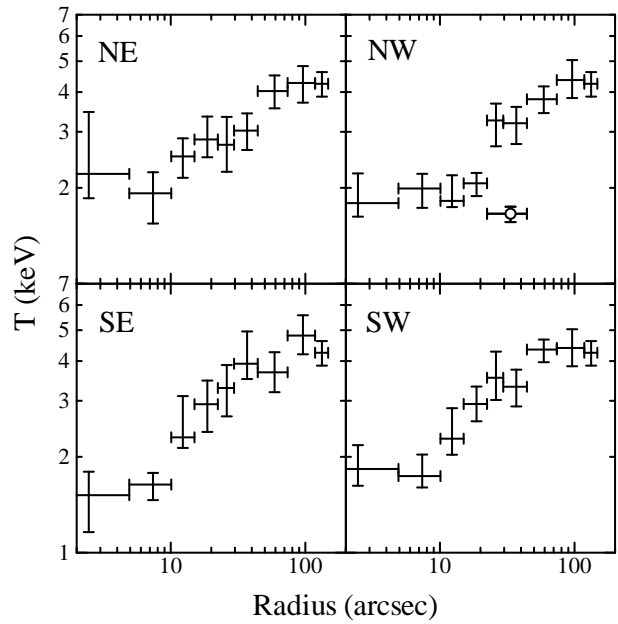} \caption{Temperature as a function of radius for the
four sectors. For the NW sector, the temperatures do not include the
tongue; the temperature of the tongue is shown separately by an open
circle.
The error bars are 90\% confidence intervals.
\label{fig:temp}}
\end{figure}

\begin{figure}\epsscale{0.60}
\plotone{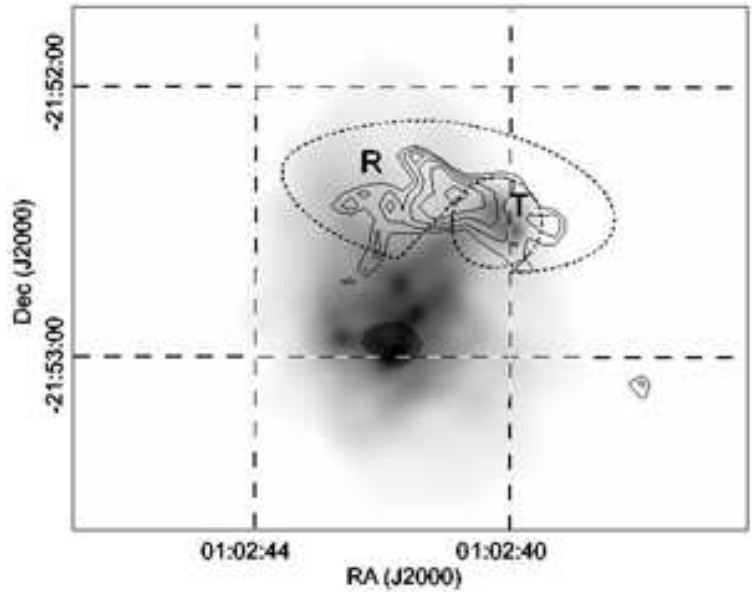} \caption{Regions for spectral analysis.
The area is the same as Figure~\protect\ref{fig:point} but the
radio relic and tongue regions are shown by dotted curves and
labeled R and T, respectively. \label{fig:relic}}
\end{figure}

\begin{figure}\epsscale{1.00}
\plottwo{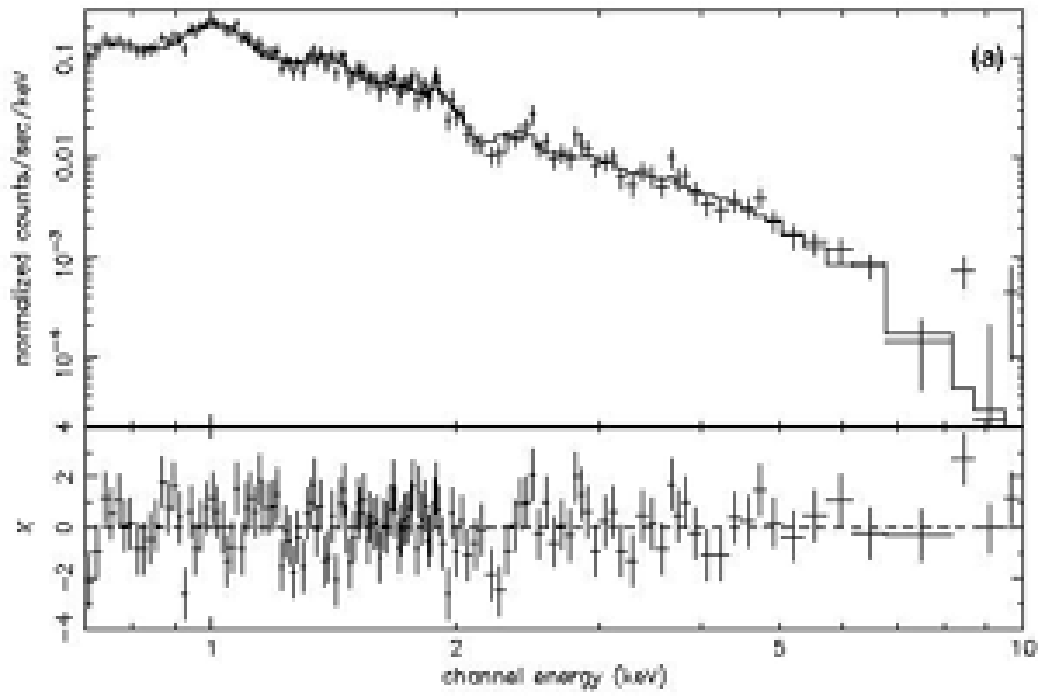}{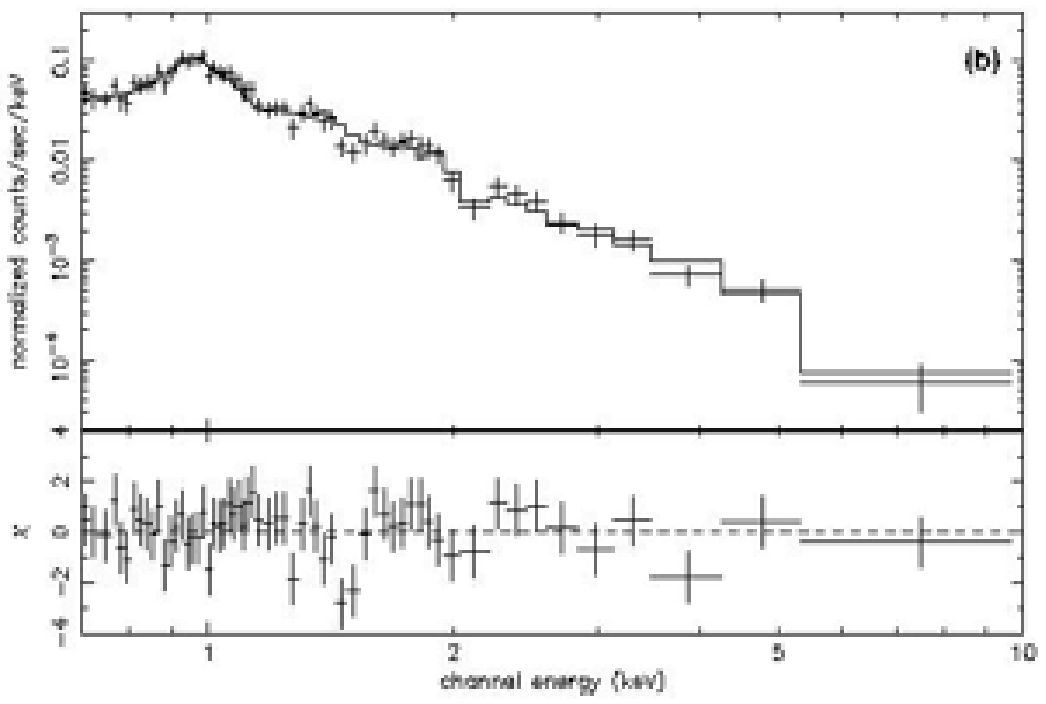}
\caption{(a) 
The upper panel shows the X-ray spectral data and best-fit
1TPL model for the radio relic region, while the lower panel plots the
residuals divided by the 1-$\sigma$ errors.
(b) Same as (a), but the spectrum is for the X-ray tongue region, and
the best-fit model is 2T$^\prime$.
\label{fig:sp}}
\end{figure}

\begin{figure}\epsscale{0.60}
\plotone{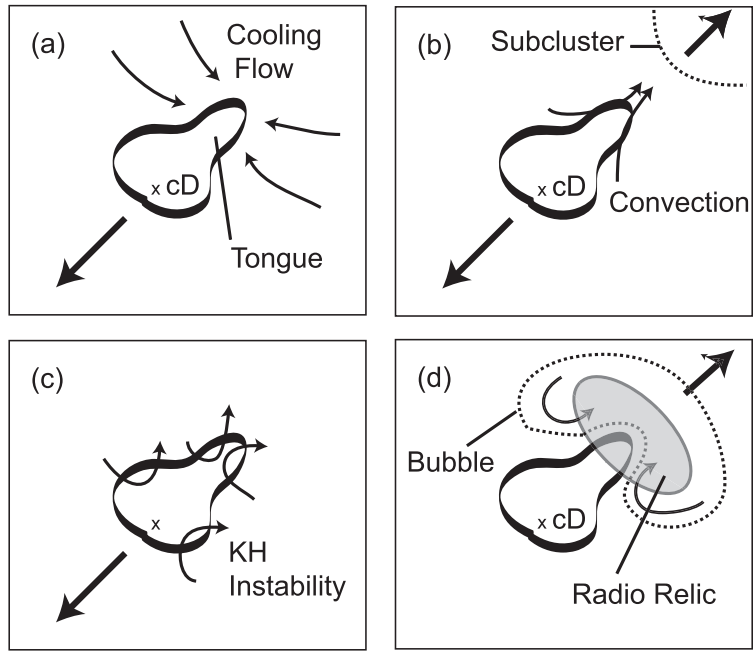} \caption{Schematic drawing of possible origins of the
 tongue: (a) a cooling wake, (b) a cluster merger, (c) Kelvin-Helmholtz
 instabilities, and (d) a buoyant radio bubble. \label{fig:ori}}
\end{figure}

In Table~\ref{tab:KH}, we compare the values of $U_{\rm KH}$ and $t_{\rm
KH}=2\pi/\omega$ for the cold fronts in Abell~133, Abell~2142,
Abell~3667, RX~J1720.1+2638, MS~1455.0+2232 and 1E0657$-$56.  The
observed parameters $T_2$, $r_2$, and $D$ used to derive the KH
instability parameters were taken from \citet{mar00}, \citet{vik01b},
and \citet{maz01a,maz02c} and \citet{mar02}.  Because \citet{sun02} do
not give the density contrast $D$ for Abell~2256 due to the complicated
geometry of the substructure, we do not discuss its cold front in the
following. We assume that the cD galaxy of Abell~133 is moving toward
the SE.  However, there is no clear large density jump in that direction
(Figure~\ref{fig:density}), and we assume that the density jump was
disrupted by the KH instability. As an upper limit, we assume the
density jump prior to disruption is less than the change in density from
$r \approx 10\arcsec$ to 30\arcsec, which is the region over which the
temperature rises rapidly. However, there is a small jump in the density
$D \approx 1.3$ at a radius of $r \approx 30\arcsec$
(Figure~\ref{fig:density}).  If we instead adopt this value of $D$, then
the maximum velocity to avoid disruption by the KH instability is
reduced to $U_{\rm KH}\sim 200\rm\; km\; s^{-1}$.  For Abell~2142,
Abell~3667, and RX~J1720.1+2638, we also present the velocities of the
cold cores, $U_{\rm obs}$, obtained from the stagnation condition at the
cold front \citep{mar00,vik01b,maz01a}. We used $U_{\rm obs}$ to derive
$t_{\rm KH}$. For Abell~133, we derived $t_{\rm KH}$ by assuming that
the core velocity is larger than the upper limit on $U_{\rm KH}$. For
MS~1455.0+2232, \citet{maz02c} do not present $U_{\rm obs}$ because of
large temperature uncertainties. Thus, we derived $t_{\rm KH}$ by
assuming $U<U_{\rm KH}$.

The small value of $U_{\rm KH}$ for Abell~133
shows that the cold core is more vulnerable to the KH instability than
in Abell~2142, Abell~3667, RX~J1720.1+2638, and MS~1455.0+2232.
This may show that there is more variety in the depths of the potentials
of substructures around cD galaxies than might be expected from the
similarity in their optical properties.  Note that Table~\ref{tab:KH}
suggests that gravity cannot completely suppress the KH instability
on large scales even for Abell~2142 and Abell~3667 ($U_{\rm obs}>U_{\rm
KH}$).  Magnetic fields may have aided in the suppression of large scale
as well as small scale instabilities in these cases
\citep[see][]{vik01a}.  Alternatively, if the cores started to move very
recently, the KH instability may have just started to develop
\citep*{maz02b}, and the cores might not have been disrupted yet.  In
Table~\ref{tab:KH}, we also present the values for 1E0657$-$56, which
was recently observed by \citet{mar02}. For this cluster, $t_{\rm KH}$
is extremely low because of the high velocity, $U_{\rm obs}$. In fact,
most of the cold gas around the substructure appears to be removed
\citep{mar02}. Abell~133 seems to be intermediate between 1E0657$-$56
and the other four clusters in the level of KH instability and in the
size of the cold front.

To assess the degree of KH instability more quantitatively, we calculate
the ratio $t_{\rm cross}/t_{KH}$, where $t_{\rm cross}$ is the crossing
time of the cool core in the cluster \citep{maz02b}. The larger $t_{\rm
cross}/t_{KH}$ is, the more vulnerable the core is to the KH
instability.  We estimate the crossing time $t_{\rm cross}$ from the
core velocity and the maximum cluster radius that a core can reach,
$r_{\rm max}$.  Let the current radius of the core be $r$, and let
$r_{200}$ be the cluster radius within which the average mass density is
200 times the critical density of the Universe ($\rho_{cr}$):
\begin{equation}
 r_{200} \equiv \left(\frac{3M_{200}}{800\pi\rho_{\rm cr}}\right)^{1/3}\:,
\end{equation}
where $M_{200}$ is the mass of the cluster within $r_{200}$.
For this estimate of $t_{\rm cross}$, we will assume that the core has a
radial orbit.  Then, the maximum radius $r_{\rm max}$ of the core is
given by
\begin{equation}
\label{eq:kin}
 (1/2)U_{\rm obs}^2 = \Phi( r_{\rm max}) - \Phi( r ) \;,
\end{equation} 
where $\Phi( r )$ is the cluster gravitational potential at $r$.  If a
cluster has an isothermal density distribution (density $\propto
r^{-2}$), the potential difference is given by
\begin{equation}
\label{eq:pot}
\Phi( r_{\rm max} ) - \Phi( r ) =
\frac{G M_{200}}{r_{200}}
 \ln\frac{r_{\rm max}}{r} \; .
\end{equation}
 From equations (\ref{eq:kin}) and (\ref{eq:pot}), we obtain
\begin{equation}
 r_{\rm max} = r \exp
\left(\frac{r_{200}U_{\rm obs}^2}{2GM_{200}}\right)\:.
\end{equation}
The cluster mass $M_{200}$ can be derived from cluster temperature $T_1$
using an empirical relation. \citet{rei01} found the relation of
\begin{equation}
 \log\left(\frac{1.4 M_{\rm 200}}{M_{\sun}}\right)
=13.735+1.710\log\left(\frac{T_1}{\rm keV}\right) 
\end{equation}
 from the X-ray data of 88 clusters.
The observed cool cores appear to be near the cluster centers.
Of course, we do not know where in the orbit the core started, or
how many times it has passed near the cluster center.
Thus, we crudely estimate the crossing times as
$t_{\rm cross}=r_{\rm max}/U_{\rm obs}$.

In Table~\ref{tab:KH}, we present the values of $T_1$, $r_{\rm max}$,
$t_{\rm cross}$, and $t_{\rm cross}/t_{\rm KH}$.  We assume $r = 0.2
r_{200}$ for all the clusters.  For Abell~133, we assume that $U_{\rm
obs}$ is larger than the upper limit on $U_{\rm KH}$, while for
MS~1455.0+2232 we assume that $U_{\rm obs} < U_{\rm KH}$.
Table~\ref{tab:KH} shows that the value of $t_{\rm cross}/t_{\rm KH}$
for Abell~133 is intermediate between that for 1E0657$-$56 and those for
the other four clusters.  This supports our argument about the relative
level of KH instability in these clusters.

Ram pressure from the ambient gas is also expected to affect the cold
gas around a cD galaxy. \citet{sar01} indicated that ram pressure would
affect the cold gas distribution at a radius which satisfies
\begin{equation}
\label{eq:ram}
 P_{\rm ram}\gtrsim P_{\rm cool}(r)\:,
\end{equation}
where $P_{\rm cool}$ is the static pressure of the gas around the cD
galaxy. Since the cD galaxy is not at the geometrical center of
Abell~133 (Figure~\ref{fig:point}), we expect that the relation is
satisfied even at $r\sim 0$. Figures~\ref{fig:density}
and~\ref{fig:temp} show that the pressure in the central region of the
cluster is $P_{\rm cool}(0)\sim 1.1\times 10^{-11}\rm\; dyn\; cm^{-2}$.
Since there is no clear cold front for this cluster, we take the gas
density around the cluster center, $n_e=0.05\rm\; cm^{-3}$, as the
density of the gas streaming relative to the cD galaxy
(Figure~\ref{fig:density}). In this case, the relative velocity of the
cD galaxy to the surrounding gas must be larger than about $400\rm\;
km\; s^{-1}$ to satisfy the relation (\ref{eq:ram}) at $r\sim 0$.  This
velocity is very similar to $U_{\rm KH}$.  We note that the criteria for
those two mechanisms, the KH instability and ram pressure, should be
similar because both involve the competition between the momentum in the
flow and gravitational acceleration.

Since the X-ray substructure around the cD galaxy of Abell~133 is
fragile as is shown by the large $t_{\rm cross}/t_{\rm KH}$, it is
unlikely that it is a merged subcluster that has survived a violent
merger.  As proposed for RX~J1720+2638 and MS~1455.0+2232, an
alternative scenario which may explain the presence of the moving
substructure is that it may be the result of the collapse of two
different perturbations in the primordial density field on two different
linear scales at nearly the same location in space
\citep{maz01a,maz02c}.  As the density perturbations grew, the smaller
scale perturbation collapsed first and formed the substructure around
the cD galaxy.  The larger scale perturbation collapsed more recently,
and forms the main body of Abell~133. The substructure around the cD
galaxy was initially located slightly offset from the center of the
cluster and it is now oscillating around the minimum of the cluster
potential well. Since the initial position of the substructure is near
to the cluster center, it is likely that the velocity is subsonic.
Recently, \citet{fuj02} showed quantitatively that substructures tend to
be located near the cluster centers.

\citet{maz01a} estimated the formation redshift of the central subclump
in RX~J1720+2638 using a relation between the formation redshift and
the temperature of a cluster. However, \citet{fuj99} showed that this
relation has a large dispersion, because for a given spatial scale the
amplitude of initial perturbations can vary significantly.  Even with
this caveat, the low temperature of the gas around the cD galaxy of
Abell~133 ($1.4$~keV) suggests a large formation redshift for the
substructure ($z \gtrsim 1$) if the virial radius was smaller than
0.5~Mpc \citep[Figure~5b in][]{fuj00}.

Since the cD galaxy is located to the SE of the radio relic and it seems
to be moving to the SE, it might be responsible for the relic.
\citet{sle01} estimated that the age of the relic is $t_{\rm radio}=
4.9\times 10^7$~yr. If the cD galaxy were located at the position of
the relic when the relic was formed, the velocity of the galaxy must be
$\gtrsim 700\;\rm km\; s^{-1}$ in the plane of the sky. This velocity is
larger than $U_{\rm KH}$ (Table~\ref{tab:KH}).  On the other hand, the
relative velocity between the cD galaxy and the whole cluster is
$220\;\rm km\; s^{-1}$ along the line of sight \citep{way97}.  This
would require that the cD velocity be at an angle of $<17\degr$ with
respect to the plane of the sky.  The value of $t_{\rm radio}$ may be
somewhat larger than $4.9\times 10^7$~yr, the transverse velocity may be
somewhat smaller than 700 km s$^{-1}$, and/or the radio source may have
been formed at some distance from the center of the cD galaxy, as is
generally true of radio lobes.  On the other hand, if the radio relic
itself is moving to the NW with the velocity of $\sim 500-600\;\rm km\;
s^{-1}$ (Mach number $0.6-0.7$) as a result of buoyancy
(\S~\ref{sec:disc_tongue_bubble}), the KH model may be more easily
reconciled with the radio relic age.

\subsubsection{Uplifted Gas by a Buoyant Bubble}
\label{sec:disc_tongue_bubble}

Recent {\it Chandra} X-ray observations
\citep[e.g.,][]{fab00,mcn00,bla01,maz02a} and earlier {\it ROSAT} images
\citep[e.g.,][]{boe93} show that radio sources at the centers of several
clusters are inflating bubbles in the ICM, which presumably consist of
relativistic particles, magnetic fields, and possibly extremely hot
thermal gas of low density.  The radio bubbles have low X-ray surface
brightnesses, presumably since the radio plasma has displaced the
ICM. These hot bubbles are buoyant and should be moving outward in the
cluster gravitational potential.  Recent numerical simulations show that
these hot bubbles may uplift cooler gas from the cluster center
\citep*{chu01,qui01,bru01a}. The nonthermal plasma in the bubble can
produce radio emission, while the uplifted dense, cool thermal gas will
emit strongly in the X-ray band.  The predicted configuration of a radio
emitting bubble and uplifted X-ray gas is very similar to the one
observed for Abell~133 (Figure~\ref{fig:point}) and also in the Virgo
cluster \citep{chu01}.  Thus, the X-ray and radio observations may
suggest that a bubble including the radio relic is uplifting the cold
tongue from the cluster center.  The deficit in X-ray surface brightness
at the radio relic can also be explained by this picture
(Figure~\ref{fig:surf}). Moreover, the fact that the tongue seems to be
in pressure equilibrium with the ambient medium is also favorable to
this model \citep{chu01,bru01a}.  However, assuming the tongue results
from the buoyant motion of the bubble, there should be another
umbrella-like X-ray excess emission at the head of the bubble due to
compression of cool gas ahead of the bubble, if the buoyant motion is in
the plane of the sky \citep[see the upper right figure of Figure~8
in][]{chu01}. Since the emission is not observed, the tongue and the
bubble may not be in the plane of the sky as is shown in the bottom
right figure of Figure~8 in \citet{chu01}.  For example, if the buoyant
motion were at an angle $45^{\circ}$ with respect to the plane of the
sky and the bubble's depth along the line of sight is the same as its
width ($\sim 50$~kpc), the observed decrease in surface brightness
($\sim 30$~\%; Figure~\ref{fig:surf}) would be consistent with the idea
that the bubble is deficient in X-ray-emitting gas.

If we assume the age of the relic is $t_{\rm radio}=4.9\times 10^7$~yr
\citep{sle01}, and the bubble including the relic was formed around the
cD galaxy, the relative upward velocity must be $v_{\rm bub}>700\;\rm
km\; s^{-1}$. More realistically, if the velocity vector of the relic is
at $45^{\circ}$ to the plane of the sky and the relic was formed by jet
activity away from the cD galaxy, say at a distance of $\sim 10$~kpc
 from the cD, the upward velocity is still $v_{\rm bub}\sim 700\;\rm km\;
s^{-1}$.  The sound velocity of the surrounding gas ($T\sim 3$~keV) is
$\sim 900\;\rm km\; s^{-1}$.  Thus, the Mach number of the rising bubble
is similar to the values of ${\cal M} \sim 0.6-0.7$ predicted by
\citet{chu01}.

Assuming that the density of the tip of the tongue is $0.03\;\rm
cm^{-3}$ and the temperature is 1.3~keV (\S~\ref{sec:spectra_reg}), the
radiative cooling time is $4\times 10^8$~yr.  This is about an order of
magnitude longer than the estimated age of the radio relic.  Thus, if
the tongue was uplifted by the buoyant motion of the radio relic, the
low temperature of the tongue cannot be the result of radiative cooling
at its current position.  The gas in the tongue presumably cooled
radiatively when the gas was located closer to the center of the cD
before it was uplifted.  In addition, adiabatic cooling may have lowered
the temperature further as the gas was lifted to regions with a lower
ambient gas pressure.

Given the high density of gas in the tongue and its large volume, one
concern is whether the radio bubble could lift such a large mass and
remain buoyant.  For the purpose of this estimate, let us assume that
the radio relic is filled with nonthermal and possibly thermal plasma
having a very low density so that its X-ray emissivity is low.
detection.  We further assume that the extent of the radio bubble along
the line of sight is similar to its largest diameter in the plane of the
sky; we treat the radio bubble as an oblate spheroid.  This is
consistent with the flattened, ``mushroom cloud'' shape expected for a
buoyant bubble \citep{chu01}.  Then, the total volume of the radio
bubble is $V_{\rm bub} \approx 4.3 \times 10^4$ kpc$^{3}$.  On the other
hand, we assume that the extent of the tongue along the line of sight is
similar to its narrower dimension; we treat it as a prolate spheroid,
which is consistent with the uplifted trail below a convective bubble
\citep{chu01}.  Then, the volume of the tongue is $V_{\rm ton} \approx
1.8 \times 10^3 $ kpc$^{3}$.  For an average tongue density of
$0.03\;\rm cm^{-3}$, this implies a total mass of $M_{\rm ton}\sim
2\times 10^9 M_{\sun}$, Thus, the average density of the material in the
radio bubble and tongue combined is $\sim$0.001 cm$^{-3}$.  This is
still much lower than the typical gas densities at similar distances on
either side of the cD galaxy in Abell~133 (Figure~\ref{fig:density}).
Thus, it is at least plausible that the radio bubble could have lifted
the tongue via buoyancy.

It is interesting to estimate the impact of the radio bubble on the
cooling flow of Abell~133. If bubbles transport a large amount of cold
gas from the cluster centers to the periphery of cooling flows, they may
explain the lack of cooled gas in the cooling flows
\citep{pet01,kaa01,tam01}.  As noted above, the total mass of the tongue
is $M_{\rm ton}\sim 2\times 10^9 \, M_{\sun}$.  This material has been
lifted over the lifetime of the radio relic, which implies an average
outflow rate of $\dot{M}_{\rm ton} \sim 40 \, M_{\sun}$ yr$^{-1}$.  This
is somewhat smaller than the cooling inflow rate derived from the X-ray
spectrum on the same radial scale (Table~\ref{tab:cool}).  However, the
total cooling radius for Abell~133 is much larger than the size of the
X-ray tongue \citep{whi97}.  Thus, the tongue seems to reside well
within the cooling radius of the cluster, and would therefore not
actually transport gas outside of the cooling radius.  Moreover, as is
shown in Table~\ref{tab:cool}, the mass deposition rate derived by
spectral analysis is increasing outward at the position of the tongue.
So, buoyant mass transport may affect the local mass cooling inflow rate
near the center of Abell~133, but it probably does not affect the
overall cooling rate.

The energy associated with the expansion and motion of the radio bubble
might affect the energetics of the X-ray gas, and might partially
counteract radiative cooling \citep{boe02}.  Assuming that the radio
bubble is adiabatic, the work done by the radio bubble on the
surrounding gas is
\begin{equation}
 W_{\rm bub}=\int_{V_0}^{V_1}P\; dV
=\int_{V_0}^{V_1}P_0 V_0^{\gamma}\; \frac{dV}{V^{\gamma}}
=\frac{P_0 V_0-P_1 V_1}{\gamma-1}\:,
\end{equation}
where $P$ and $V$ are the pressure and the volume of the radio bubble,
respectively, and $\gamma$ is the adiabatic constant. The subscripts 0
and 1 stand for the values when the radio bubble is formed and those at
present, respectively. If we assume that the radio bubble was formed
near the cluster center and has been nearly in pressure equilibrium with
the surrounding gas, $P_0=1.7\times 10^{-10}$ and $P_1=7.9\times
10^{-11}\;\rm dyn\; cm^{-2}$ (Figures~\ref{fig:density}
and~\ref{fig:temp}). Moreover, if we assume that the volume of the
observed radio bubble is $V_1=4.3\times 10^4\rm\; kpc^3$ and
$\gamma=5/3$, the initial volume is $V_0=2.7\times 10^4\rm\; kpc^3$ and
the work is $W_{\rm bub}=5.2\times 10^{58}\rm\; erg$.  Moreover, the
gravitational energy released by matter displacement is
\begin{equation}
 W_{\rm dis}\approx \frac{V_1+V_0}{2}\rho_{\rm hot}
g_{\rm ton}l_{\rm ton}\;,
\end{equation}
where $\rho_{\rm hot}$ is the density of the gas surrounding the tongue,
and $l_{\rm ton}$ is the length of the tongue. The gravitational
acceleration around the tongue is given by
\begin{equation}
 g_{\rm ton}\approx \frac{P_0-P_1}{\rho_{\rm hot}l_{\rm ton}}\;.
\end{equation}
Thus, the gravitational energy is $W_{\rm dis}\approx 9.1\times
10^{58}\rm\; erg$, and the total energy released by the bubble motion is
$W=W_{\rm bub}+W_{\rm dis}\approx 1.4\times 10^{59}\rm\; erg$. If the
work has been done during the past $4.9\times 10^7$~yr, the rate of work
is $9\times 10^{43}\rm\; erg\; s^{-1}$. On the other hand, the
luminosity of the cooling flow component is $2.5\times 10^{43}\rm\;
erg\; s^{-1}$ (see \S~\ref{sec:spectra_cool}), which is smaller than
$W$. Thus, the energy dissipated by the expansion of the radio bubble
may compensate the energy radiated by cooling gas at least for this
cluster at present, although it is uncertain what fraction of the
dissipated energy is converted to thermal energy \citep[but
see][]{qui01}.

Although the buoyant radio bubble model may explain the observed tongue,
at present it is not clear what conditions are required to produce
uplifting of dense cool gas.  For example, the numerical simulations of
\citet{bru02} do not show the formation of a cold X-ray tongue.

\subsection{The Radio Relic}
\label{sec:disc_relic}

\subsubsection{Origin of the Radio Relic}
\label{sec:disc_relic_origin}

Assuming that the unusual radio source to the north of the cluster
center in Abell~133 is related to the cluster, there have been at least
two suggestions as to its origin.  First, the source might be a cluster
radio relic \citep[e.g.,][]{sle01}.  Radio relics are diffuse radio
sources with steep spectra which are not associated with a radio galaxy
in the cluster and which are generally located far from the center of
the cluster \citep[e.g.,][]{rot97}.  Generally speaking, radio relic
sources tend to be moderately polarized.  This and their outer location
distinguish radio relics from radio halos.  Radio relics (and radio
halos) are generally associated with clusters which appear to be
undergoing a merger.  A common theoretical interpretation is that radio
relics are produced by acceleration, re-acceleration, or adiabatic
compression of relativistic electrons and magnetic fields by merger
shocks.

\begin{figure}\epsscale{0.7}
\plotone{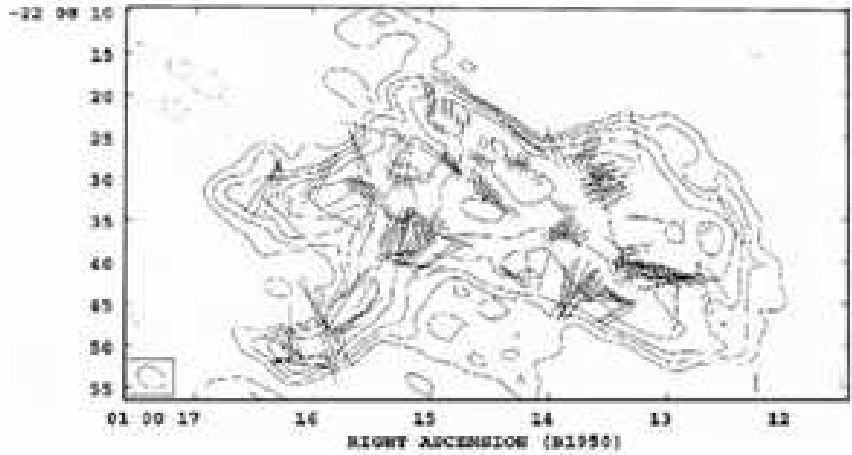}
\caption{Radio polarization of the radio relic source at 1.465 GHz 
based on data described in \citet{sle01}.
The lines are parallel to the electric vector, and
the length is proportional to the fractional polarization.
The line in the rectangle at the lower right corresponds to a polarization
of 10\%.
The contours show the radio surface brightness;
the positive contours are (91, 182, 304, 607, 1214, and 2732) $\mu$Jy
beam$^{-1}$.
\label{fig:radio_polar}}
\end{figure}

Some of the radio properties of the source in Abell~133 seem at variance
with those of other cluster radio relics.  First, the radio source is
projected only $\approx$ 40 kpc from the cD galaxy of Abell~133.  Of
course, it is possible that it is actually at a much larger radial
distance, and only appears near the center in projection.  There are
several other radio relics seen in projection near cluster centers,
including the sources in Abell~13 and Abell~4038 \citep{sle01}.  Second,
the integrated polarization of the radio relic source in Abell~133 is
low \citep[$\sim$ 2.3\%][]{sle01}, unlike most radio relics.  However,
the low integrated polarization is mainly the result of variations in
the direction of the polarization across the source.
Figure~\ref{fig:radio_polar} shows the radio polarization at 1.465 GHz
as a function of position in the radio relic source; the data comes from
the same observations discussed in \citet{sle01}.  The data in
Figure~\ref{fig:radio_polar} are not corrected for Faraday rotation.  In
a number of comparatively large areas within the main body of the relic
the polarized fraction varies between 3 and 20 percent, the larger
fractions suggesting that either the Faraday rotation in these areas is
not large or that it is fairly uniform.  Note that even rotation
measures as small as $RM \ga 20$ rad m$^{-2}$ would produce Faraday
rotations of $\ga 45^\circ$ at 1.4 GHz.  In general, radio sources
located near the centers of cluster cooling flows show very high Faraday
rotations \citep*[$RM \ga 800$ rad m$^{-2}$; e.g.,][]{tbg94,geo94}. The
fact that strong polarization is observed in the Abell~133 relic without
correction for Faraday rotation may indicate the relic is located in
front of, rather than within or behind, the cluster cool core.

If the radio relic were produced by a merger shock and the magnetic
field were strongly affected by compression in the shock, one would
expect the magnetic field direction (perpendicular to the electric
vector) to be mainly parallel to the shock front.  Indeed, we see in
Figure~\ref{fig:radio_polar} that the electric vectors closest to the
concave northern edge of the relic have the required orientation to
suggest that this may be a shock front with a tangential magnetic field
direction.  Of course, any other mechanism which produced a compression
of the northern edge of the radio source might have the same effect; if
the radio source has risen buoyantly to the north, one might also expect
some compression of this edge.  However, this must be an uncertain
interpretation until the electric vectors can be corrected for Faraday
rotation.

Several other features in Figure~\ref{fig:radio_polar} are of interest.
The polarization near the bright knots at the tip of the X-ray tongue 
($01^{\rm h}00^{\rm m}13\fs5$; $-22\degr08\arcmin40\arcsec$, B1950) and
further down the tongue to the SE is quite strong and particularly
variable in direction.  Perhaps this indicates that the relic lies
behind the tongue, and that the emission from those regions of the relic
suffers strong and variable Faraday rotation by the comparatively cool
gas in the tongue.

In the two filaments of radio emission extending to the east and
south-east, the polarizations are mainly perpendicular to the filaments,
which implies that the magnetic field is parallel to them.  The origin
of these filaments may be connected to the magnetic field lines
\citep[e.g.,][]{ens01a,ens02}.
The radio relic in Abell~133 is rather filamentary.  Although there are
several possible origins for this structure, \citet{ens01a} and
\citet{ens02} suggested that the structure results from the compression
of old radio plasma by merger shocks.

On the other hand, our {\it Chandra} observations are largely
inconsistent with a merger shock origin for the radio relic.  First of
all, no evidence for a shock is seen in the image
(Figure~\ref{fig:point}), X-ray colors (Figure~\ref{fig:color}), or
spectra (Figure~\ref{fig:sp}) of the radio relic region.  If the radio
relic coincided with a shock, this region should be unusually bright in
X-rays, due mainly to compression of the cluster gas in the shock.
Actually, most of the radio relic region is unusually faint in X-rays
(Figure~\ref{fig:surf}).  The exception is the region where the X-ray
tongue overlaps the radio relic.  However, the tongue is due to cool
thermal gas, and is thus not a result of a merger shock.

An alternative explanation of the radio relic is that it is an old radio
lobe from the central radio source associated with the cD galaxy in
Abell~133 \citep{riz00}.  An early radio map of the source suggested a
bridge of emission connecting the cD radio source with the radio relic
\citep{riz00}, but this does not appear in a more sensitive, higher
resolution image \citep{sle01}.  Many of the radio properties of the
relic are consistent with its interpretation as an old radio lobe.
Similar radio lobes associated with cD galaxies in cooling flow clusters
generally have rather steep radio spectra \citep[e.g.,][]{geo94,riz00}.
The displacement of the radio relic from the central cD might be due to
buoyancy (\S~\ref{sec:disc_tongue_bubble}), or to the motion of the
central cD, perhaps as a result of a subcluster motion
(\S~\ref{sec:disc_tongue_KH}). One difficulty is that one usually finds
two radio lobes, presumably produced by a double-sided jet.  However, if
the central cD is moving to the southeast, the two bubbles would be
displaced in the same direction and might either overlap in projection
or have merged together.

Our {\it Chandra} observations are probably more consistent with the
interpretation of this radio source as a relic radio lobe formerly
energized by the central cD, rather than a merger-shock generated
cluster radio relic.  Radio lobes associated with cD galaxies at the
centers of cooling flow clusters produce holes in the X-ray emission
\citep[e.g.,][]{fab00,mcn00,bla01}, suggesting that the radio plasma has
displaced the thermal, X-ray emitting gas.  In Abell~133, the radio
relic region is also somewhat fainter in X-rays than the surrounding
region (Figure~\ref{fig:surf}).  As noted in
\S~\ref{sec:disc_tongue_bubble}, many properties of the X-ray tongue are
consistent with gas which has been uplifted by a buoyant radio bubble.
In fact, the overall radio and X-ray properties of the relic source in
Abell~133 are similar to the eastern radio bubble (the ``ear'') of M87
in the Virgo cluster \citep*{owe00}.  This mushroom-shaped radio bubble
has a shape similar to the radio source in Abell~133, and also has a
very steep radio spectrum.  The M87 ear is filamentary, although mainly
around its edge.  In M87, there is a linear extension of X-ray emission
due to cool, thermal gas which connects the center of the cluster with
the radio bubble \citep{boe95}, similar to the tongue in Abell~133.  The
radio and X-ray structure of M87 have been interpreted as the result of
a buoyant radio bubble \citep*{chu01,bru02,you02}. One difference
between M87 and Abell~133 is that there is radio emission connecting the
ear in M87 with the nucleus of the galaxy, while the radio relic in
Abell~133 does not show any strong radio bridge with the central cD.
This might be a result of a lower level of current AGN activity in the
cD galaxy in Abell~133 compared to M87.

In summary, our X-ray observations are more consistent with the
interpretation of the radio relic in Abell~133 as a displaced radio
lobe from the central cD, rather than as a merger shock-induced relic.

\subsubsection{Magnetic Field in the Radio Relic}
\label{sec:disc_relic_B}

In \S~\ref{sec:spectra_reg}, we found an (at best) marginal detection of
nonthermal X-ray emission from a radio relic.  We now use these spectral
results to constrain the magnetic field in the radio relic, under the
assumption that this nonthermal emission is due to inverse Compton (IC)
scattering.  In the model with a power-law component (Model~1TPL in
\S~\ref{sec:spectra_reg}), the X-ray flux of the power-law component is
$2.2\times 10^{-13}\rm\; erg\;cm^{-2}\; s^{-1}$ for $2-10$~keV and the
photon index is $\Gamma = 1.7^{+0.3}_{-1.1}$.  The radio synchrotron
flux density of the relic is 137~mJy and 
flux density spectral index is $\alpha = 2.1\pm 0.1$ (flux density
proportional to $\nu^{-\alpha}$), both at $1.425$~GHz \citep{sle01}.
For emission from a single power-law electron energy distribution,
the IC and synchrotron emission have the same spectral index.
This would require $\Gamma = \alpha + 1$, which is not consistent with
our measured values.
However,
the IC X-ray emission in the {\it Chandra} band would be due to
relatively low energy relativistic electrons, which would produce very
low frequency radio emission.  Both observations and models suggest that
the radio spectrum is considerably flatter at lower frequencies
\citep{sle01}, and thus the X-ray and radio spectral indices might be
consistent.  We will adopt the X-ray spectral index and compare the
X-ray flux with the radio flux density of $35.5 \pm 4.3 $ Jy at 80 MHz,
the lowest frequency at which the flux density is determined accurately.
This radio flux density includes the cD galaxy and another nearby radio
source, but they are much weaker than the radio relic.  Using the
standard expressions for the IC and synchrotron emission assuming a
power-law electron energy distribution \citep[e.g.,][]{sar86}, we find
that the required magnetic field in the relic is $B \approx 1.2\rm\;\mu
G$.  This is about one order of magnitude smaller than the magnetic
field \citep{sle01} derived from models for the radio spectral index ($B
\approx 10\rm\;\mu G$) or minimum energy arguments applied to the radio
flux ($B \approx 14\rm\;\mu G$).  This might indicate that the magnetic
field is filamentary and highly variable, which may also be suggested by
the radio image of the the relic.  However, given the marginal detection
of the nonthermal X-ray emission and the disparity between the magnetic
field values required by the radio and X-ray observations, we believe it
is best to treat the X-ray spectral result as an upper limit of the
nonthermal flux.  This implies a lower limit on the radio relic magnetic
field of $B \ge 1.2\rm\;\mu G$, which is consistent with the radio
results but not constraining.

\section{Conclusions}
\label{sec:conc}

We have presented the results of a {\it Chandra} observation of the
cluster of galaxies Abell~133. The X-ray image shows a tongue of
emission extending from the cD galaxy to the NW; the tongue extends in
the direction of and partially overlaps the radio relic. There is no
optical counterpart of the tongue. The X-ray surface brightness at the
position of the radio relic is smaller than the surrounding region
except for the tongue, suggesting that the radio plasma has displaced
the thermal gas in this region.  The X-ray spectrum indicates that the
emission from the tongue is thermal, and the temperature is
distinctly lower than that of the ambient hot gas.

We discussed several possible origins for the tongue; two of these seem
reasonably consistent with the X-ray and radio data.  First, the tongue
may be the result of a Kelvin-Helmholtz instability acting on the cool
core surrounding the cD galaxy as the cD moves through surrounding,
lower density gas.  We show that the cD galaxy and the cold gas around
it must be moving with a velocity of $\gtrsim 400$ km s$^{-1}$ for a
large scale KH instability to develop between the cold core and the
ambient hot intracluster medium.  The critical velocity for the KH
instability is several times smaller for Abell~133 than for other
clusters which show smooth, well-defined cold fronts.  The required
motion of the cD galaxy is consistent with its offset position in the
cluster relative to the X-ray centroid determined at large radii.  This
offset can be explained if the velocity of the cD galaxy is $\gtrsim
400\;\rm km\;^{-1}$.  Thus the onset of a KH instability is at least
plausible.

The second possibility is that the radio relic is a buoyant radio
bubble, and that the tongue was uplifted by the motion of this bubble.
Recent numerical simulations indicate that buoyant bubbles of nonthermal
plasma produced by AGN activity can uplift cold gas from the cluster
center \citep{chu01,qui01,bru02}. For Abell~133, we suggest that the
bubble is the radio relic, and that it uplifts the X-ray tongue.  The
velocity of the radio bubble expected from the age of the radio relic is
consistent with that predicted numerically. The energy dissipated by the
moving bubble may affect the cooling flow in Abell~133.

The origin of the filamentary radio relic source in Abell~133 remains
uncertain.  It may be a radio source that is generated by a shock wave
from merging clusters, similar to the relics which are often seen in the
outer regions of clusters.  In this case, the radio emitting electrons
would have been accelerated, re-accelerated, and/or compressed by the
cluster merger shock.  Alternatively, the radio relic might be a
displaced radio lobe (or lobes) from the radio source associated with
the central cD galaxy in the cluster.  The lack of evidence for a merger
shock near the relic, the anti-correlation between the radio and X-ray
emission across most of the relic, and the association of the relic with
the tongue of cool, dense X-ray gas all are more consistent with the
relic being a displaced radio lobe or bubble, rather than being a
cluster merger-shock induced radio relic.

\acknowledgments

We are grateful to E. Blanton, S. Randall, I. Tanaka, and S. Oya for
useful comments.  Support for this work was provided by the National
Aeronautics and Space Administration through $Chandra$ Award Numbers
GO1-2122X and GO1-2123X, issued by the $Chandra$ X-ray Observatory
Center, which is operated by the Smithsonian Astrophysical Observatory
for and on behalf of NASA under contract NAS8-39073.  L. R. acknowledges
support from NSF grant AST-0071167 to the University of Minnesota.
H. A. thanks CONACyT for financial support under grant 27602-E.

\clearpage

\begin{deluxetable}{cccccc}
\footnotesize
\tablecaption{Average Spectrum of Abell 133\label{tab:tot}}
\tablewidth{0pt}
\tablehead{
\colhead{} & \colhead{$T$} 
& \colhead{$Z$}   & \colhead{$N_{\rm H}$}   &
\colhead{$F(2-10\rm\; keV)$} &  \colhead{$\chi^2$/dof} \\
\colhead{Observatory}  & \colhead{(keV)} & 
\colhead{($Z_{\sun}$)}     & \colhead{($10^{20}\rm\; cm^{-2}$)}  &
\colhead{($10^{-11}\rm erg\; cm^{-2}\; s^{-1}$)} &  
}
\startdata
Chandra  &$3.4^{+0.0}_{-0.1}$
                          &$0.59^{+0.04}_{-0.04}$ 
                                         &$2.92^{+0.69}_{-0.63}$&0.75 
                                         & 452.8/315 \\
Chandra\tablenotemark{a}  
        &$1.4^{+0.1}_{-0.3}, 3.7^{+0.2}_{-0.2}$
                          &$0.59^{+0.06}_{-0.05}$ 
                                         &$3.8^{+0.8}_{-0.9}$&0.75
                                         & 347.1/313 \\
Einstein &$4.0^{+2.9}_{-1.4}$& 0.5\tablenotemark{b}
                                      &1.58\tablenotemark{c} &1.31 
                                         & \nodata \\
EXOSAT   &$3.8^{+2.0}_{-0.9}$& $<1.1$
                                      & $0.28^{+0.99}_{-0.26}$ &1.43
                                         & 24.60/17 \\
\enddata
\tablenotetext{a}{
The ratio of the emission measure of the cold component to that of the
hot component
is $0.083 \pm 0.035$.}
\tablenotetext{b}{Assumed}
\tablenotetext{c}{Assumed  (Galactic absorption)}
\end{deluxetable}

\begin{deluxetable}{cccccccc}
\footnotesize
\tablecaption{Spectra for the X-ray Tongue, Radio Relic, and cD Center 
Regions
\label{tab:region}}
\tablewidth{0pt}
\tablehead{\colhead{} &
\colhead{} & \colhead{$T_1$} & \colhead{$T_2$}
& \colhead{$Z$} & \colhead{$\Gamma$}   
& \colhead{$N_{\rm H}$} & \colhead{$\chi^2$/dof}  \\
\colhead{Region}  &
\colhead{Model}  & \colhead{(keV)} & \colhead{(keV)} & 
\colhead{($Z_{\sun}$)}     & \colhead{}  &
\colhead{($10^{20}\rm\; cm^{-2}$)} & \colhead{}
}
\startdata
Relic & 1T   &$2.7_{-0.2}^{+0.2}$&\nodata 
                  &$0.62_{-0.12}^{+0.15}$&\nodata 
                       &$3.6_{-1.5}^{+2.1}$ & 156.0/128 \\
      & 1TPL &$2.2_{-0.2}^{+0.4}$
                  &\nodata & 0.62\tablenotemark{a} 
                                           &$1.7_{-1.1}^{+0.3}$ 
                                           &$4.8_{-3.3}^{+4.0}$ 
                                                 & 147.2/127 \\
      & 2T   &$1.2_{-0.4}^{+0.9}$&$2.9_{-1.7}^{+\infty}$ 
                           &$0.66_{-0.30}^{+0.0}$ &\nodata 
                                           &$4.2_{-2.8}^{+3.0}$ 
                                                 & 151.1/126 \\
Tongue& 1T   &$1.4_{-0.0}^{+0.3}$&\nodata 
                           &$0.28_{-0.06}^{+0.08}$&\nodata 
                                           &$5.9_{-5.9}^{+0.0}$
                                                 & 75.8/57 \\
      & 1TPL &$1.4_{-0.1}^{+0.0}$&\nodata &0.28\tablenotemark{a} 
                           &$0.9_{-1.4}^{+0.6}$ &$3.8_{-3.8}^{+5.0}$
                                                 & 59.1/56 \\
      & 2T   &$1.1_{-0.2}^{+0.2}$&$2.3_{-0.3}^{+1.2}$ 
                           &$0.77_{-0.24}^{+0.67}$ &\nodata 
                                           &$3.1_{-3.1}^{+5.1}$
                                                 & 52.6/55 \\
     &2T$'$&$1.3_{-0.1}^{+0.0}$& 3.2\tablenotemark{b}
                        &$0.66_{-0.25}^{+0.55}$ (0.72\tablenotemark{b} ) 
                                           &\nodata & 
                                               1.3\tablenotemark{b} 
                                                        & 54.6/58 \\
cD Center&1T   &$1.7_{-0.1}^{+0.1}$& \nodata
                        &$0.42_{-0.11}^{+0.15}$
                                           &\nodata 
                                           &$5.4_{-4.3}^{+4.2}$
                                                        &62.4/58 \\

\enddata

\tablenotetext{a}{Fixed at the value of 1T model.}
\tablenotetext{b}{Fixed at the values for the ambient gas.}

\end{deluxetable}


\begin{deluxetable}{cccccc}
\footnotesize
\tablecaption{Cooling Flow Spectral Fits \label{tab:cool}}
\tablewidth{0pt}
\tablehead{
\colhead{$r$} & \colhead{$T_{\rm High}$} 
& \colhead{$Z$}   & \colhead{$\Delta N_{\rm H}$}   &
\colhead{$\dot{M}$} & \colhead{$\chi^2$/dof}  \\
\colhead{(arcsec)}  & \colhead{(keV)} & 
\colhead{($Z_{\sun}$)}     & \colhead{($10^{20}\rm\; cm^{-3}$)}  &
\colhead{($M_{\sun}\rm\; yr^{-1}$)}  & 
}
\startdata
$0-27$    &$2.7_{-0.1}^{+0.1} $
             &$1.0_{-0.1}^{+0.2} $
                &$34_{-5}^{+4} $ 
                   &$56_{-15}^{+11} $ 
                      & 195.7/154  \\
          &$(2.7_{-0.1}^{+0.1})$
             &$(0.85_{-0.11}^{+0.12})$
                &$(0_{-0}^{+12})$
                   &$(19_{-4}^{+11})$
                      & (164.1/140) \\
$27-45$   &$3.4_{-0.3}^{+0.2} $
             &$0.78_{-0.07}^{+0.20} $ 
                &$28_{-7}^{+6} $  
                   &$22_{-7}^{+9} $ 
                      & 178.2/136  \\
          &$(3.4_{-0.3}^{+0.2})$
             &$(0.68_{-0.12}^{+0.14})$
                &$(0_{-0}^{+15})$
                   &$(9_{-4}^{+6})$
                      & (154.5/127) \\
$45-150$  &$4.7_{-0.5}^{+0.6} $ 
             &$0.34_{-0.11}^{+0.11} $ 
                &$25_{-9}^{+35} $  
                   &$8_{-2}^{+6} $ 
                      & 227.5/224  \\
          &$(4.7_{-1.3}^{+0.3})$
             &$(0.35_{-0.12}^{+0.15})$
                &$(29_{-29}^{+>100}) \tablenotemark{a}$
                   &$(8_{-8}^{+13})$
                      & (215.3/210)
\enddata
\tablecomments{Parenthesis show the values when the photon data 
below 0.9~keV are ignored.}
\tablenotetext{a}{Cannot be constrained.}
\end{deluxetable}


\begin{deluxetable}{ccccccccccc}
\tabletypesize{\scriptsize}
\tablecaption{Parameters for the KH Instability \label{tab:KH}}
\tablewidth{0pt}
\tablehead{
\colhead{} & \colhead{$T_2$} & \colhead{$r_2$}
& \colhead{$D$}   & \colhead{$U_{\rm KH}$}   
& \colhead{$U_{\rm obs}$} & \colhead{$t_{\rm KH}$}
& \colhead{$T_1$} & \colhead{$r_{\rm max}$} 
& \colhead{$t_{\rm cross}$} & \colhead{$t_{\rm cross}/t_{\rm KH}$}
\\
\colhead{Clusters}  & \colhead{(keV)} & \colhead{(kpc)} &
\colhead{}     & \colhead{($\rm km\; s^{-1}$)}  &
\colhead{($\rm km\; s^{-1}$)} & \colhead{($10^8$~yr)}
& \colhead{(keV)} & \colhead{(Mpc)} & \colhead{($10^8$~yr)}
& \colhead{} 
}
\startdata
Abell~133     & 1.4 &  25 & $<2.5$ & $<380$ & \nodata & 
                                           $<1.4$\tablenotemark{a}
              & 4.0\tablenotemark{c} & $>0.33$\tablenotemark{a} 
              & $>8.4$\tablenotemark{a} & $>6.0$\tablenotemark{a} \\
Abell~2142    &   7 & 250 &    2   & 720    &  900    & 5.8    
                  & 9.4\tablenotemark{c} & 0.57 & 6.2 & 1.1 \\ 
Abell~3667    &   4 & 400 &    3.9 & 850    & 1400    & 6.9   
                  & 6.5\tablenotemark{c} & 0.66 & 4.6 & 0.67 \\ 
RX~J1720.1+2638
              &   6 & 200 &    2   & 670    &  500    & 8.3     
                  & 5.6\tablenotemark{d} & 0.40 & 7.8 & 0.94 \\ 
MS~1455.0+2232
              &   5 & 200 & 2.7    & 760    & \nodata & 
                                           $>5.8$\tablenotemark{b} 
              & 4.8\tablenotemark{e} & $<0.42$\tablenotemark{b} 
              & $<5.3$\tablenotemark{b} & $<0.91$\tablenotemark{b} \\ 
1E0657$-$56   &   7 &  25 &    3.8 & 1100   & 3500    & 0.2
                  &14.8\tablenotemark{f} & 2.1 & 5.9 & 30 \\ 
\enddata
\tablenotetext{a}{$U_{\rm obs}>380\rm\: km\: s^{-1}$ is assumed.}
\tablenotetext{b}{$U_{\rm obs}<760\rm\: km\: s^{-1}$ is assumed.}
\tablenotetext{c}{\citet{dav93}}
\tablenotetext{d}{\citet{maz01a}}
\tablenotetext{e}{\citet{maz02c}}
\tablenotetext{f}{\citet{mar02}}

\end{deluxetable}

\end{document}